\documentstyle[12pt,titlepage]{article}

\renewcommand{\theequation}{\thesection.\arabic{equation}}
\font\tenrm=cmr10
\def\fl{\flushleft}

\def\L{{\cal L}}

\def\a{\alpha}
\def\b{\beta}
\def\d{\delta}
\def\e{\epsilon}

\def\g{\gamma}
\def\l{\lambda}
\def\n{\eta}

\def\z{\zeta}

\def\O{{\cal O}}

\def\s{\sigma}
\def\k{\kappa}
\def\t{\theta}
\def\vphi{\varphi}
\def\w{\omega}

\def\hf{\frac{1}{2}}
\def\der{\partial}
\def\bq{\begin{equation}}
\def\eq{\end{equation}}
\def\brr{\begin{eqnarray}}
\def\err{\end{eqnarray}}
\def\ba{\left(\begin{array}}
\def\ea{\end{array}\right)}
\def\pp{\hbox{\ooalign{$\displaystyle\int$\cr$-$}}}
\def\ppa{\hbox{\ooalign{$-$\cr$\displaystyle\int_{-L/2}^{+L/2}$}}}
\def\ppb{\hbox{\ooalign{$-$\cr$\displaystyle\int_{-\infty}^{+\infty}$}}}

\def\ba{\left(\begin{array}}
\def\ea{\end{array}\right)}
\def\gbig {\hbox{\Large\it g}}
\input epsf.tex
\epsfverbosetrue
\begin{document}
\pagestyle{empty}
\begin{flushright}
CERN-TH. 7017/93
\end{flushright}
\begin{large}
\begin{center}{\bf EFFECTIVE $d=2$ SUPERSYMMETRIC
               LAGRANGIANS}\end{center}
\vspace{-.3in}
\begin{center} {\bf FROM $d=1$ SUPERMATRIX MODELS} \end{center}
\end{large}
\begin{center}
    RAM BRUSTEIN , MICHAEL FAUX$^{*)}$ and BURT A. OVRUT$^{*),**)}$
    \end{center}
\vspace{-.3in}
\begin{center} Theory Division, CERN \end{center}
\vspace{-.3in}
\begin{center} CH-1211 Geneva 23, Switzerland  \end{center}
\begin{center} ABSTRACT \end{center}

We discuss $d=1, {\cal N}=2$ supersymmetric matrix models
and exhibit the associated $d=2$ collective field theory in the
limit of dense eigenvalues.  From this theory we construct,
by the addition of several new fields, a $d=2$ supersymmetric
effective field theory, which reduces to the collective field
theory when the new fields are replaced with their vacuum
expectation values.  This effective theory is Poincare invariant
and contains perturbative and non-perturbative information
about the associated superstrings. We exhibit instanton solutions
corresponding to the motion of single eigenvalues and discuss their
possible role in supersymmetry breaking.

\vspace*{1.8cm}
\noindent
\rule[.1in]{16.5cm}{.002in}\\
\noindent
$^{*)}$ Work supported in part by DOE under Contract No.
DOE-AC02-76-ERO-3071.\\
$^{**)}$ On sabbatical leave from the Department of Physics, University
of Pennsylvania,
Philadelphia, PA 19104, U.S.A.\\
\vspace*{0.5cm}

\begin{flushleft} CERN-TH.7017/93 \\
September 1993
\end{flushleft}
\vfill\eject

\setcounter{page}{1}
\pagestyle{plain}

\renewcommand{\theequation}{1.\arabic{equation}}
\setcounter{equation}{0}
{\fl{\bf 1. Introduction}}

Non-perturbative interactions in string theory are believed to
determine a number of important quantities, such as the strength of
supersymmetry breaking \cite{dixon}.  Matrix models \cite{done}
offer a unique opportunity to learn about non-perturbative aspects of
string theory.  The $d=1$ matrix
models have related to them string
theories with a low number of degrees of freedom, propagating in
1+1 space-time dimensions \cite{jevrev}.  These matrix models have the
power to describe non-perturbative phenomena in the associated
string theories.  Moreover, there are indications that some
non-perturbative features are common to all string
theories \cite{shenker}.
By studying the generic features of non-perturbative behavior
in 1+1 dimensional string theories, one may therefore
learn about more realistic string theories, such as those in four
dimensions.

To use $d=1$ matrix models for the purpose of understanding
non-perturbative effects in string theory, it is essential
to first construct the complete
two-dimensional effective Lagrangian for the associated $d=2$
string theory. Once that is achieved, one can look
for non-perturbative phenomena, such as instantons. This entire
program has already been carried out \cite{bo}, with
positive results, in the case of $d=1$ bosonic matrix models. In
this case, the nonperturbative effective Lagrangian
of the strings was constructed and its
fundamental symmetry, a non-compact shift symmetry,
$\z\rightarrow\z+c$, in one
of its bosonic fields $\z$, was shown to be
broken by instantons in a single
eigenvalue of the matrix model.  This was done using the methods of
collective field theory \cite{das}.  A notable property of the collective
field theory is the presence of a space-dependent coupling
parameter.  This has been consistently interpreted as deriving
from a field dependent coupling in an effective theory
which reduces to the collective field theory when the field in question
attains a space-dependent vacuum expectation value.
In this paper we extend this previous result to the
supersymmetric case.  We choose the simplest interesting
construction, which involves a $d=1, \cal{N}=2$ supersymmetric
matrix model \cite{marpar}. The simpler $d=1, \cal{N}=1$ matrix model is not of
interest since it is a non-interacting theory.

To arrive at a collective field description of a matrix model,
it is necessary to first isolate the sub-theory of
the matrix eigenvalues.
The process of extracting the eigenvalue theory from the supermatrix
model and then correctly identifying a canonical collective
field description is rather involved.
This has previously been attempted by several groups
\cite{dabh,supjev,sa,jd,jf}
with only partial success.  The first half
of this paper is devoted to a careful analysis of this problem.
A feature of the collective field theories obtained in this manner
is that they are not Poincare invariant and
not supersymmetric.  A related feature
is the existence of a coupling parameter in the collective
field theory which is  space-dependent. The second half of this paper is
devoted
to describing, reconciling, and interpreting these facts.  We interpret the
non-Poincare invariant, non-supersymmetric collective field theory as deriving
from a particular Poincare invariant, supersymmetric effective theory  when
certain ``heavy"
fields in the effective theory are frozen in their vacuum expectation
values (VEV's).

An important question is which $d=2$ supersymmetry the effective theory
should have. We demonstrate that, of all  $d=2, (p,q)$ supersymmetries, it is
$(1,1)$ supersymmetry which appropriately relates to the $d=1, {\cal{N}}=2$
supersymmetic matrix model.
The construction of a $d=2$ supersymmetric effective superstring
Lagrangian using matrix models can be accomplished in several
related ways.  An important consistency check
relevant to the work described in
this paper results  from the demonstration that each of these
ways yield the same results.
We illustrate the various possibilities in figure 1.

\ \vspace{1pc}

\centerline{\tenrm Figure~1. Different ways to obtain a supersymmetric
            effective Lagrangian using matrix models}
The boxes in this figure represent intermediate steps in the
construction of the supersymmetric effective theory.  The upper
line represents the transformation from a bosonic matrix model
through various associated bosonic theories.  The steps
labeled 1-3 thus respectively represent the extraction of the
bosonic eigenvalue theory through a suitable restriction of the
Hilbert space, the construction of the bosonic collective field theory
from the eigenvalue theory, and finally the identification of a
bosonic effective theory which is Poincare invariant and which reduces
to the bosonic collective field theory when a ``heavy" field is
frozen in its VEV.  The bottom line represents
an analagous derivation starting from a supersymmetric matrix model.
Thus, the steps labeled $I-III$ respectively represent
the extraction of a supersymmetric eigenvalue theory by a suitable
restriction of the Hilbert space, the construction of the
associated collective field theory, and finally the identification of
an effective field theory which is both Poincare invariant and
supersymmetric and which reduces to the collective field theory
when ``heavy" fields are replaced with their VEV's.  The steps
labeled $A-D$ represent direct supersymmetrizations of the
various bosonic theories.  Thus, a supersymmetric effective
theory can be obtained by supersymmetrizing the bosonic effective
theory.  Alternatively, any intermediate bosonic theory could
be supersymmetrized and then the remaining steps in the bottom
line could be followed until a supersymmetric effective theory
is obtained.  In this paper we emphasize the route,
labeled in figure 1 by $A-I-II-III$,
from a bosonic
matrix model to a supersymmetric effective theory.
However, at every step we check that
our result is, in fact, the appropriate supersymmetrization of the
corresponding bosonic theory.  We thus show that the diagram
indicated in figure 1 commutes completely.

The paper is structured as follows.

In section 2 we discuss the bosonic matrix model.  We define the theory,
quantize it, and explain how a quantum mechanics of matrix eigenvalues
can be extracted from the theory upon suitable restriction of the
Hilbert space.

In section 3 we discuss the supersymmetric analog of the bosonic
matrix model.  We introduce a $d=1, {\cal{N}}=2$ supersymmetric
matrix model, quantize the theory and show how a supersymmetric
quantum mechanics can be extracted from the theory upon suitable
restriction of the Hilbert space.  Not suprisingly, this is more
subtle than the bosonic case.

In section 4 we represent the supersymmetric quantum mechanics,
extracted from the supersymmetric matrix model in section 3, in
terms of collective fields.  This is done by introducing a new spatial
parameter $x$, which is a continuous extension of the discrete eigenvalue
index.  The collective fields aggregate the distinct matrix
eigenvalues into fields defined over $x$ and $t$.
We show that the large $N$ limit can be taken in two
distinct ways and that these two ways can be taken independently
over any regions of $x$.  In the first of these, the high density case,
the eigenvalues ``pack" densely over $x$.  The high density
collective fields defined in this way are ordinary two dimensional
fields.  In the second case, only a finite number of eigenvalues
populate the associated region of $x$. In this section we present the
mathematical details of the derivation of the high density
collective field theory
related to the $d=1, {\cal{N}}=2$
supersymmetric matrix model.  We discuss
subtleties of regularization and canonicalization of the theory.
We show that the supersymmetric collective field theory possesses a
single coupling parameter which blows up at finite points in space.
We end this section by exhibiting the high density collective
field theory, with canonically normalized fields. As mentioned above,
the theory is neither Poincare invariant nor supersymmetric.

In section 5 we derive a (1,1) supersymmetric effective theory which
reduces to the collective field theory derived in section 4 when certain
fields are frozen in their VEV's.  We discuss two dimensional
$(p,q)$ supersymmetry and demonstrate that only (1,1) supersymmetry is
compatible with the collective field theory.  The
Poincare invariant, (1,1) supersymmetric effective field
Lagrangian derived in this section is the essential result of this
paper.

In section 6 we solve the Euclidean equations of motion and find solutions
 corresponding to the motion of individual eigenvalues in the low density
regions. We exhibit, explicitly, the eigenvalue instantons alluded to
above.  We then briefly describe how we expect these instantons
to break the supersymmetry of the effective theory.

\renewcommand{\theequation}{2.\arabic{equation}}
\setcounter{equation}{0}
{\fl{\bf 2. The Bosonic Matrix Model}}

In this section we briefly review the bosonic matrix model.  We begin with a
description of the classical theory and its symmetries. We then quantize the
theory and show that an effective quantum theory involving only the matrix
eigenvalues can be constructed provided the Hilbert space is suitably
restricted. The results of this section are  known. We discuss them here in
order to motivate the extension to the supersymmetric case and also to set our
notation.

The fundamental variable in the bosonic matrix model is a time-dependent
$N\times N$ Hermitian matrix, $M(t)$.  Its dynamics are described by
the Lagrangian,
\bq L(\dot{M},M)=\hf Tr{\dot{M}^2}-V(M).
 \label{lagclassmat} \eq
The potential is taken to be polynomial,
\bq V(M)=\sum_na_n Tr{M^n},
 \eq
where the $a_n$ are real coupling parameters.  The mass dimension of
$M$ is $-\hf$ so that the $a_n$ have positive mass dimension
$(n+2)/2$.  The momentum conjugate to $M$ is the $N\times N$ Hermitian
matrix $\Pi_M(t)=\dot{M}$.  It follows that the associated Hamiltonian
is given by
\bq H(\Pi_M,M)=\hf Tr{\Pi_M^2}+V(M).
 \eq
The matrix $M$ remains Hermitian under the transformation
$M\rightarrow
{\cal{U}}^{\dagger}M{\cal{U}}$ where ${\cal{U}}$
is an arbitrary $N\times N$ complex
matrix.  The Lagrangian is invariant under such
a transformation provided it is global and that
${\cal{U}}\in U(N)$.  Thus the
classical theory possesses a global $U(N)$ symmetry.
We proceed to quantize this theory.
As stated above, it is of great interest
to extract an effective quantum theory of matrix eigenvalues.
This procedure is complicated
by the fact that it is necessary to suitably restrict the Hilbert
space of states in order to diagonalize the matrix momentum operator
$\hat{\Pi}_M$.
This poses a difficulty when attempting to express the
quantum theory using path integral language, which is the natural
language for a discussion of the nonperturbative issues which are
our main concern.  This complication is subtle and the extraction
of the effective eigenvalue theory using path integrals from the outset,
although possible, is relatively complicated. Such procedures are
explained at various places in the literature, e.g. \cite{sakita,bo}.
An equivalent procedure is to first extract the relevant eigenvalue theory
using
canonical operator quantum mechanics.  The passage to a path integral
description is then straightforward.
We proceed with a
description of this method. A detailed discussion of the following
calculation is given in Appendix A.

In the $M$ basis, the operator $\hat{\Pi}_M$, constructed to
satisfy $[\hat{\Pi}_{M_{ij}},\hat{M}_{kl}]=-i\d_{ik}\d_{jl}$, is given
by
\bq \hat{\Pi}_{M_{ij}}=-i\frac{\der}{\der{M}_{ij}}.
 \eq
Thus, the quantum operator Hamiltonian is
\bq \hat{H}=-\hf\sum_{ij}\frac{\der}{\der{M}_{ij}}
                \frac{\der}{\der{M}_{ji}}+V({M}).
 \eq
Since $M$ is Hermitian, there exists, at every time $t$,
a unitary matrix, $U(t)$, such that
$M=U^\dagger\l U$, where $\l$ is a time dependent
diagonal matrix consisting of the
eigenvalues of $M$.  This is a useful parameterization of $M$.
The operator
${\der}/{\der{M}}$ can then be decomposed into a sum of
operators involving ${\der}/{\der{\l}}$ and ${\der}/{\der{U}}$.
Then, by restricting attention
to only those states $|s>$
which are annihilated by
$\der/\der{U}$, the $U(N)$ ``singlet" sector of the Hilbert space,
it can be shown that
$\hat{H}|s>=\hat{H}_s|s>$, where
\bq \hat{H}_s=\sum_i\Biggl\{-\hf\frac{\der^2}{\der{\l}_i^2}
    -\sum_{j\ne i}\frac{1}{{\l}_i-{\l}_j}
    \frac{\der}{\der{\l_i}}+V({\l}_i)\Biggr\}.
 \eq
Since $\hat{\Pi}_{\l_i}=i{\der}/{\der{\l}_i}$ is the
momentum operator conjugate to ${\hat{\l}}_i$ this effective
Hamiltonian can be expressed as
\bq \hat{H}_s=\sum_i\Biggl\{\hf\hat{\Pi}_{\l_i}^2
    -i\sum_{j\ne i}\frac{1}{{\l}_i-{\l}_j}\hat{\Pi}_{\l_i}
    +V({\l}_i)\Biggr\}.
 \eq
It is then straightforward, using well known techniques, to show
that the quantum mechanics of the singlet sector is governed by the
following partition function,
\bq Z_N(a_n)=\int[d\l]\exp{i\int dt L_s(\dot{\l},\l)},
 \eq
where
\bq L_s(\dot{\l},\l)=\sum_i\Biggl\{\hf\dot{\l}_i^2-V_{eff}(\l_i)\Biggr\}
 \label{lags} \eq
and
\bq V_{eff}(\l_i)=V(\l_i)+\hf(\sum_{j\ne i}\frac{1}{\l_j-\l_i})^2.
 \eq
Once again, a detailed derivation of this result is given in
Appendix A.
The Lagrangian $L_s$ given in (\ref{lags})
is the appropriate Lagrangian for
studying the dynamics of the $U(N)$ singlet sector of the
bosonic quantum
mechanical matrix model.
\renewcommand{\theequation}{3.\arabic{equation}}
\setcounter{equation}{0}
{\fl{\bf 3. A Supersymmetric Matrix Model}}

In this section we present a supersymmetric matrix model.  After
introducing the classical theory and its symmetries, we then quantize
the theory and show that the effective
quantum theory of the matrix eigenvalues reduces to a
supersymmetric quantum mechanics provided the Hilbert space is suitably
restricted.  This particular model was originally presented by
Marinari and Parisi\cite{marpar}.
We present a brief discussion of it here as it
is essential to the main results in this paper.
The reduction to supersymmetric quantum
mechanics has also been presented elsewhere\cite{dabh},
but we present an
alternative method which we find  illuminating.

There are  many $d=1$ supersymmetries enumerated by the
number of supersymmetric charges, ${\cal{N}}$. The simplest nontrivial
supersymmetric extension of the bosonic theory presented in the last section
involves a $d=1, {\cal{N}}=2$ supersymmetry.  This is because $d=1,
{\cal{N}}=1$
supersymmetry does not admit interactions.  We extend the bosonic
theory by letting the fundamental variable be a time-dependent
$N\times N$ matrix whose elements are $d=1, {\cal{N}}=2$ complex
superfields. We further restrict this matrix to be Hermitian.
The reason for choosing Hermitian matrices rather than real symmetric
matrices is the following.  Matrix models involving real symmetric
matrices generate triangulations of string worldsheets which are both
orientable and non-orientable whereas Hermitian matrix models
describe only orientable worldsheets.  Since we want our matrix
model to describe a two dimensional supersymmetric string theory,
we must assume the existence of supersymmetry on the associated
string worldsheet.  The worldsheet is thus a spin manifold and a
spin manifold is necessarily orientable.  This motivates the choice of
Hermitian matrices.  We wish to point out that the complex $d=1,
\cal{N}=2$ supermultiplets are reducible under supersymmetry.
Regardless of this fact, the supersymmetric quantum mechanics of the
matrix eigenvalues, which we will extract from the matrix model,
will involve only real irreducible $d=1, \cal{N}=2$ multiplets,
since the diagonal elements of a Hermitian matrix are real.
We now present the details of the classical theory.

As just described,
the fundamental variable of the supermatrix model is a
time-dependent, $N\times N, d=1, {\cal{N}=2}$ Hermitian matrix superfield,
\bq \Phi_{ij}=M_{ij}(t)+i\t_1\Psi_{1ij}(t)+i\t_2\Psi_{2ij}
    +i\t_1\t_2F_{ij}(t),
 \eq
where $\t_1$ and $\t_2$ are real anticommuting parameters,
$M_{ij}$ and $F_{ij}$ are $N\times N$ bosonic Hermitian matrices and
$\Psi_{1ij}$ and $\Psi_{2ij}$ are $N\times N$
fermionic Hermitian matrices.
We note that $(\t\Psi_{ij})^\dagger=\Psi_{ij}^\dagger\t=-\t\Psi_{ji}^*$.
Thus, $\Phi_{ij}=\Phi_{ij}^\dagger$.  The Lagrangian  is
\bq L=\int d\t_1d\t_2\Biggl\{\hf Tr D_1\Phi D_2\Phi+iW(\Phi)\Biggr\},
 \eq
where the superpotential, $W$, is  a polynomial in $\Phi$,
\bq W(\Phi)=\sum_n b_n Tr\Phi^n,
 \eq
$b_n$ are real coupling parameters,
and $D_I$ are superspace derivatives,
\bq D_I=\frac{\der}{\der\t_I}+i\t_I\frac{\der}{\der t}
 \eq
for $I=1,2$.
In component fields, the Lagrangian reads
\brr L &=& \sum_{ij}\Biggl\{\hf(\dot{M}_{ij}\dot{M}_{ji}+F_{ij}F_{ji})
     +\frac{\der W(M)}{\der M_{ij}}F_{ij}\Biggr\} \nonumber \\
     & & -\frac{i}{2}\sum_{ij} (\Psi_{1ij}\dot{\Psi}_{1ji}
     +\Psi_{2ij}\dot{\Psi}_{2ji})
     -i\sum_{ijkl}\Psi_{1ij}\frac{\der^2W(M)}{\der M_{ij}\der M_{kl}}
     \Psi_{2kl}.
 \label{lagmatsusy} \err
The $d=1, {\cal{N}=2}$ supersymmetry
transformation law on the component matrices is given by
\brr \d M_{ij} &=& i\n^1\Psi_{1ij}+i\n^2\Psi_{2ij} \nonumber \\
     \d\Psi_{1ij} &=& \n^1\dot{M}_{ij}+\n^2F_{ij} \nonumber \\
     \d\Psi_{2ij} &=& \n^2\dot{M}_{ij}-\n^1F_{ij} \nonumber \\
     \d F_{ij} &=& i\n^2\dot{\Psi}_{1ij}-i\n^1\dot{\Psi}_{2ij},
 \err
where $\n^1$ and $\n^2$ are anticommuting constants.
It is straightforward to check that
this is a symmetry of the Lagrangian (\ref{lagmatsusy}).
The momenta conjugate to the matrices $M, \Psi_1$ and $\Psi_2$ are
\brr \Pi_{M_{ij}} &=& \dot{M}_{ji} \nonumber \\
     \Pi_{\Psi_{1ij}} &=& -\frac{i}{2}\Psi_{1ij} \nonumber \\
     \Pi_{\Psi_{2ij}} &=& -\frac{i}{2}\Psi_{2ij}.
 \err
Thus, the Hamiltonian is given by
\bq  H = \sum_{ij}\Biggl\{\hf\Pi_{M_{ij}}\Pi_{M_{ji}}
     -F_{ij}F_{ji}-\frac{\der W(M)}{\der M_{ij}}F_{ij}\Biggr\}
     +i\sum_{ijkl}\Psi_{1ij}\frac{\der^2W(M)}
     {\der M_{ij}\der M_{kl}}\Psi_{2kl}.
 \label{hammatsusy} \eq
Note that $H$ does not depend on $\Pi_{\Psi_{Iij}}$ for $I=1,2$.
Now, $\Phi_{ij}$ remains a Hermitian matrix of superfields under
the transformation $\Phi\rightarrow{\cal{U}}^\dagger\Phi{\cal{U}}$ where
${\cal{U}}$ is an arbitrary $N\times N$ matrix of complex numbers.
The Lagrangian is invariant under such a transformation provided
that ${\cal{U}}\in U(N)$.  Thus the classical theory possesses a global
$U(N)$ symmetry.

Before quantizing the theory, we eliminate the
auxiliary matrix $F_{ij}$.  Its equation of motion reads
\bq F_{ij}=-\frac{\der W(M)}{\der M_{ji}}.
 \eq
Eliminating $F_{ij}$ with this equation, the Lagrangian
then becomes
\brr L &=& \sum_{ij}\Biggl\{\hf\dot{M}_{ij}\dot{M}_{ji}
     -\hf\frac{\der W(M)}{\der M_{ij}}\frac{\der W(M)}{\der M_{ji}}\Biggr\}
     \nonumber \\
     & & -\frac{i}{2}\sum_{ij}(\Psi_{1ij}\dot{\Psi}_{1ji}
     +\Psi_{2ij}\dot{\Psi}_{2ji})
     -i\sum_{ijkl}\Psi_{1ij}\frac{\der^2W(M)}{\der M_{ij}\der M_{kl}}
     \Psi_{2kl}.
 \label{lagmatsusynof} \err
This is symmetric with respect to the nonlinear $d=1, {\cal{N}=2}$
supersymmetry transformation,
\brr \d M_{ij} &=& i\n^1\Psi_{1ij}+i\n^2\Psi_{2ij} \nonumber \\
     \d\Psi_{1ij} &=& \n^1\dot{M}_{ij}-\n^2
     \frac{\der W(M)}{\der M_{ji}} \nonumber \\
     \d\Psi_{2ij} &=& \n^2\dot{M}_{ij}+\n^1
     \frac{\der W(M)}{\der M_{ji}}.
 \err
and also with respect to the $U(N)$ transformation
\brr M &\rightarrow& {\cal{U}}^\dagger M{\cal{U}}, \nonumber \\
    \Psi_1 &\rightarrow& {\cal{U}}^\dagger\Psi_1{\cal{U} } \nonumber \\
    \Psi_2 &\rightarrow& {\cal{U}}^\dagger\Psi_2{\cal{U}}.
 \err
With $F_{ij}$ eliminated, the classical Hamiltonian becomes
\bq H=\sum_{ij}\Biggl\{\hf\Pi_{M_{ij}}\Pi_{M_{ji}}
     +\hf\frac{\der W(M)}{\der M_{ij}}\frac{\der W(M)}{\der M_{ij}}\Biggr\}
     +i\sum_{ijkl}\Psi_{1ij}\Psi_{2kl}
     \frac{\der^2W(M)}{\der M_{ij}\der M_{kl}}.
 \eq
We proceed to quantize the theory.

Canonical quantization is achieved by promoting the matrices
$M, \Pi_M$ and $\Psi_I$ to operators and by imposing the following
relations,
\brr [\hat{\Pi}_{M_{ij}},\hat{M}_{kl}] &=& -i\d_{ik}\d_{jl} \nonumber \\
     \{\hat{\Psi}_{Iij},\hat{\Psi}_{Jkl}\} &=& \d_{IJ}\d_{ik}\d_{jl}.
 \err
For the fermions, it is useful to define complex operators,
\brr \hat{\Psi} &=& \frac{1}{\sqrt{2}}(\hat{\Psi}_1+i\hat{\Psi}_2)
     \nonumber \\
     \hat{\bar{\Psi}} &=& \frac{1}{\sqrt{2}}(\hat{\Psi}_1-i\hat{\Psi}_2).
 \err
It then follows that
\bq \{\hat{\bar{\Psi}}_{ij},\hat{\Psi}_{kl}\}=\d_{ik}\d_{jl}.
 \eq
We can thus choose $\hat{\Psi}$ and $\hat{\bar{\Psi}}$, respectively, to
be annihilation and creation operators for fermions.  The quantum
operator Hamiltonian can now be written
\bq \hat{H}=\sum_{ij}\Biggl\{\hf\hat{\Pi}_{M_{ij}}\hat{\Pi}_{M_{ji}}
     +\hf\frac{\der W(\hat{M})}{\der\hat{M}_{ij}}
     \frac{\der W(\hat{M})}{\der\hat{M}_{ij}}\Biggr\}
     +\hf\sum_{ijkl}[\hat{\bar{\Psi}}_{ij},\hat{\Psi}_{kl}]
     \frac{\der^2W(\hat{M})}{\der\hat{M}_{ij}\der\hat{M}_{kl}}.
 \eq
Upon appropriate restriction to a
subspace of the full Hilbert space, this theory
reduces to a supersymmetric quantum mechanics.
We proceed to show this.  A detailed derivation
of the following calculation is given in Appendix B.

We work in the $M$ basis, so that $\hat{\Pi}_M=-i\der/\der{M}$.
We then parameterize $M$ in terms of its eigenvalues and angular
variables, as discussed in section 2.  Thus, $M=U^\dagger\l U$,
where $\l$ is a diagonal matrix of time-dependant
eigenvalues and $U(t)$ is a unitary
matrix.  The operator $\der/\der{M}$ is then decomposed into a sum of
operators involving $\der/\der\l$ and $\der/\der{U}$.
We define a ``rotated" fermion matrix $\chi=U\Psi U^\dagger$.
Note that $U$ diagonalizes $M$ but that
$\chi$ is not diagonal.
It is possible to show, on states $|S>$ which are
annihilated by both $\der/\der{U}$ and by $\hat{\chi}_{ij}$, where
$i\ne j$, the $U(N)$ ``singlet" sector of the Hilbert space,
that $\hat{H}|S>=\hat{H}_S|S>$, where
\bq \hat{H}_S=\sum_i\Biggl\{\hf\hat{\Pi}_{\l_i}^2
    +i\frac{\der w}{\der\l_i}\hat{\Pi}_{\l_i}
    +\hf(\frac{\der W}{\der\l_i})^2
    +\hf\frac{\der w}{\der\l_i}\frac{\der W}{\der\l_i}\Biggr\}
    +\hf\sum_{ij}[\bar{\chi}_{i},\chi_{j}]\frac{\der^2w}{\der\l_i^2}.
 \label{hammer} \eq
In (\ref{hammer}), and henceforth, we abbreviate $\chi_{ii}$
by writing $\chi_i$,
$\hat{\Pi}_{\l_i}=-i\der/\der\l_i$, and
\bq w=-\sum_i\sum_{j\ne i}\ln|\l_i-\l_j|.
 \label{dasher} \eq
We note that $w$ has the following properties,
\bq \frac{\der w}{\der\l_i} = -\sum_{j\ne i}\frac{1}{\l_i-\l_j},
 \eq
and
\brr \frac{\der^2 w}{\der\l_m\der\l_n}=
     \left\{\begin{array}{ll}
            \sum_{k\ne i}{1}/{(\l_i-\l_k)^2} & \mbox{$;  m=n$} \\
            -1/(\l_m-\l_n)^2 & \mbox{$;m\ne n$}
      \end{array}\right. .
  \err
It is then straightforward, using well known techniques, to show that
the quantum mechanics of the singlet sector is governed by the
following partition function,
\bq Z_N(b_n)=\int[d\l][d\bar{\chi}][d\chi]
    \exp({i\int dt L_S})
 \eq
where
\bq L_S=\sum_i\Biggl\{\hf\dot{\l}_i^2-\hf(\frac{\der W_{eff}}{\der\l_i})^2
    -\frac{i}{2}(\bar{\chi}_i\dot{\chi}_i-\dot{\bar{\chi}}_i\chi_i)\Biggr\}
    -\sum_{ij}\bar{\chi}_i\chi_j
    \frac{\der^2W_{eff}}{\der\l_i\der\l_j},
  \label{lageigsusy} \eq
and
\bq W_{eff}(\l_i)=W(\l_i)+w(\l_i)
 \eq
For convenience we
rewrite this Lagrangian as follows,
\brr L &=& \sum_i\Biggl\{\hf\dot{\l}_i^2
     -\hf(\frac{\der W}{\der\l_i})^2
     -\frac{\der w}{\der\l_i}\frac{\der W}{\der\l_i}
     -\hf(\frac{\der w}{\der\l_i})^2
     -\frac{i}{2}(\bar{\chi}_i\dot{\chi}_i
     -\dot{\bar{\chi}}_i\chi_i)\Biggr\} \nonumber \\
     & & -\sum_{ij}
     \Biggl\{\frac{\der^2 W}{\der\l_i\der\l_j}\bar{\chi}_i\chi_j
     +\frac{\der^2 w}{\der\l_i\der\l_j}\bar{\chi}_i\chi_j\Biggr\}
 \label{lageig} \err
In passing from (\ref{lageigsusy}) to (\ref{lageig}) we have dropped
the subscript $S$.  It is henceforth assumed that we are describing
only the singlet sector of the matrix model.
Once again, a detailed derivation of this result
is given in
Appendix B.
\pagebreak

\renewcommand{\theequation}{4.\arabic{equation}}
\setcounter{equation}{0}
{\fl{\bf 4. Supersymmetric Collective Field Theory}}

In this section we introduce the notion of collective fields.  This
is a powerful construction which will allow us, in subsequent sections,
to investigate the physics
of the supersymmetric matrix model.
The formalism is particularly suited to studying this model in
the large $N$ limit.

We begin by introducing
a continuous real parameter, $x$, constrained to lie
in the interval $-L/2<x<L/2$.
On this line segment define
``collective fields",
\brr \vphi(x,t) &=& \sum_i\Theta(x-\l_i(t)) \nonumber \\
     \psi(x,t) &=& -\sum_i\d(x-\l_i(t))\chi_{i}(t) \nonumber \\
     \bar{\psi}(x,t) &=& -\sum_i\d(x-\l_i(t))\bar{\chi}_{i}(t).
 \label{colldef} \err
The parameter $x$ is a
continuous extension of the discrete eigenvalue index.  For finite
$N$ the two dimensional fields $\vphi, \psi$, and $\bar{\psi}$
have a finite
number of independent modes.
They are thus not ordinary unconstrained fields.
Eventually, we take the limit $N\rightarrow\infty,
L\rightarrow\infty$.  We can take this limit in one of two ways;
we may let $N/L\rightarrow$ finite or we may let
$N/L\rightarrow\infty$.  In the first case the average
density of eigenvalues over $x$ remains finite.  In this case, the
collective fields remain unwieldy as mathematical tools.  In the second
case, however, the density of eigenvalues becomes infinite.
In this case, it can be shown,
by representing the theta and delta functions by
Gaussian integrals and then taking the desired limit, that,
modulo a subtlety which we will discuss below, the collective
fields shed their constraints and become ordinary two dimensional fields.
We proceed to represent the eigenvalue
Lagrangian, (\ref{lageig}), in terms of the collective fields defined in
(\ref{colldef}). We begin by keeping both $N$ and $L$ finite.
The collective field representation of the
eigenvalue Lagrangian is then nothing more than a reparameterization.
We then take the $N\rightarrow\infty, L\rightarrow\infty$ limit.
We will see that a careful reevaluation of the significance of the
collective field Lagrangian is then warranted.

{\fl{\it 4.1 Finite N Collective Field Theory}}

Using the definitions (\ref{colldef}) it is easily seen that
\brr \int dx\frac{\dot{\vphi}^2}{2\vphi'}
     &=& \sum_i\hf\int dx\{\frac{\sum_j\d(x-\l_j)}{\vphi'(x)}\}
     \d(x-\l_i)\dot{\l}_i^2 \nonumber \\
     &=& \hf\sum_i\int dx\d(x-\l_i)\dot{\l}_i^2 \nonumber \\
     &=& \hf\sum_i\dot{\l}_i^2,
 \label{set} \err
where a dot represents a time derivative and a prime represents a
derivative with respect to $x$.
This offers an alternative
representation of the first term in the eigenvalue Lagrangian
(\ref{lageig}).  We can apply similar techniques to the terms,
\brr \hf\sum_i\dot{\l}_i^2 &=& \int dx\frac{\dot{\vphi}^2}{2\vphi'}
     \label{dd1} \\
     -\hf\sum_i(\frac{\der W}{\der\l_i})^2 &=&
     -\hf\int dx\vphi'W'(x)^2 \label{dd2} \\
     -\frac{i}{2}\sum_i\bar{\chi}_{i}\dot{\chi}_{i}
     &=& \int dx\{-\hf\frac{\bar{\psi}\dot{\psi}}{\vphi'}
     +\frac{i}{2}\frac{\dot{\vphi}}{\vphi^{'2}}\bar{\psi}\psi'\}
     \label{dd3} \\
     \sum_{ij}\frac{\der^2W}{\der\l_i\der\l_j}\bar{\chi}_i\chi_j
     &=& \int dx\frac{W''(x)}{\vphi'}\bar{\psi}\psi,
 \label{dd4} \err
where, on the right hand side,
\bq W(x)=\sum_nb_nx^n.
 \eq
The other terms in the eigenvalue Lagrangian
contain factors of $w$.  Since $w=-\sum_i\sum_{i\ne j}\ln{|\l_i-\l_j|}$,
we must properly regulate the collective field expression.
We will now discuss this is some detail.

Generically, we encounter terms of the following sort,
\brr \sum_i\sum_{j\ne i}\frac{f(\l_i,\l_j)}{\l_i-\l_j}
     &=& \sum_i\sum_{j\ne i}\int dxdy\d(x-\l_i)\d(y-\l_j)
     \frac{f(x,y)}{x-y} \nonumber \\
     &=& \sum_{ij}\pp dxdy\d(x-\l_i)\d(y-\l_j)\frac{f(x,y)}{x-y}
     \nonumber \\
     &=& \pp dxdy\vphi'(x)\vphi'(y)\frac{f(x,y)}{x-y}.
 \label{ex} \err
The symbol $\pp$ designates ``principal part" of the integral,
which is defined as follows.
Given an integral over a real function with a simple pole,
\bq \int_{-L/2}^{+L/2}dx\frac{\phi(x)}{x-y},
 \label{expr} \eq
where $\phi(x)$ is analytic and $y$ is a constant, we note that, for
$|y|<L/2$, the integral is not a-priori
well defined. The ``principle part" of the integral is defined by the
following limiting procedure,
\bq \ppa dx\frac{\phi(x)}{x-y}
    =\lim_{\e\rightarrow 0}(\int_{-L/2}^{y-\e}dx+\int_{y+\e}^{+L/2}dx)
    \frac{\phi(x)}{x-y}.
 \label{ppdef} \eq
This removes the point $x=y$ from the range of integration.
Thus, in passing from the first line in (\ref{ex}) to the second we have
shifted the regulator $j\ne i$ onto the continuous coordinate by
implicitly invoking $x\ne y$. Note that when
$N$ is finite, which in this subsection it is, equation
(\ref{ex}) is merely a series of identities.  The more complicated case
when $N\rightarrow\infty$ will be discussed in detail later.
We can now compute the
remaining terms in the collective field Lagrangian.

Using the techniques discussed above it is straightforward
to prove that
\brr \sum_i(\frac{\der w}{\der\l_i})^2
     &=& \frac{1}{3}\pp dxdydz\frac{\vphi'(x)\vphi'(y)\vphi'(z)}
         {(x-y)(x-z)}, \label{ee1} \\
     -\sum_i\frac{\der w}{\der\l_i}\frac{\der W}{\der\l_i}
     &=& \pp dxdy\frac{\vphi'(x)\vphi'(y)}{(x-y)}W'(x),
     \label{ee2} \\
     \sum_{ij}\frac{\der^2w}{\der\l_i\der\l_j}\bar{\psi}_i\psi_j
     &=& -\pp dxdy\frac{1}{(x-y)^2}
         \Biggl\{\bar{\psi}(x)\psi(y)-\frac{\vphi''(y)}{\vphi'(x)}
         \bar{\psi}(x)\psi(x)\Biggr\} \nonumber \\
     &=& \pp\frac{1}{(x-y)}\Biggl\{\bar{\psi}(x)\psi'(y)
         -\frac{\vphi''(y)}{\vphi'(x)}\bar{\psi}(x)\psi(x)\Biggr\}.
         \label{ee3}
 \err
To obtain the last line of (\ref{ee3}) we have integrated by parts.
We may now assemble the full collective field Lagrangian.
Inserting the expressions
(\ref{dd1})-(\ref{dd4}) and (\ref{ee1})-(\ref{ee3}),
into the Lagrangian (\ref{lageig}), we obtain
\brr L &=& \int dx\Biggl\{
       \frac{\dot{\vphi}^2}{2\vphi'}
       -\hf\vphi'W'(x)^2
       +\frac{W''(x)}{\vphi'}\bar{\psi}\psi \nonumber \\
       & & \hspace{.3in}
       -\frac{1}{2\vphi'}(\bar{\psi}\dot{\bar{\psi}}
       +\dot{\bar{\psi}}\bar{\psi})
       +\frac{i}{2}\frac{\dot{\vphi}}{\vphi^{'2}}
       (\bar{\psi}\psi'-\bar{\psi}'\psi)\Biggr\} \nonumber \\
       & & +\frac{1}{3}\pp dxdydz\frac{\vphi'(x)\vphi'(y)\vphi'(z)}
       {(x-y)(x-z)}, \nonumber \\
       & & +\pp dxdy\frac{\vphi'(x)\vphi'(y)}{(x-y)}W'(x)
       \nonumber \\
       & & +\pp\frac{1}{(x-y)}\Biggl\{\bar{\psi}(x)\psi'(y)
       -\frac{\vphi''(y)}{\vphi'(x)}\bar{\psi}(x)\psi(x)\Biggr\}.
 \label{lagcollpp} \err

{\fl{\it 4.2 The $N$-dependence Of The Superpotential.  }}

Before we consider the case $N\rightarrow\infty, L\rightarrow\infty$
we should first discuss a relevant issue concerning the $N$ dependance
of the superpotential.
Recall that the superpotential was expressed as a polynomial,
\bq
W(x)=\sum_n b_nx^n.
\eq
It turns out, if the coefficients $b_n$ depend on $N$ in a  specific manner,
that, when the limit $N\rightarrow\infty$ is taken, the matrix partition
function actually describes an ensemble of two dimensional super-Riemann
surfaces.  This is what
allows us to interpret the matrix models as describing
string theory.  Since our interest in matrix models is to help us
better understand string theory, we should accordingly impose that
the coefficients $b_n$ have the appropriate $N$
dependence.  The correct dependence is that $b_n$ should scale as $N^{1-n/2}$.
If we write
$b_n=\frac{1}{n!}N^{1-n/2}\tilde{c}_n$,
where the $\tilde{c}_n$ do not depend on $N$,
the superpotential becomes
\bq W(x)=\sum_n\frac{1}{n!}N^{1-n/2}\tilde{c}_nx^n.
 \eq
Since $N$ is finite, we can also make the
following shift,
\bq x\rightarrow x+\sqrt{N}\beta,
 \eq
where $\beta$ is an arbitrary real constant.
This induces a shift in the superpotential,
\bq W(x)\rightarrow\sum_n\frac{1}{n!}N^{1-n/2}c_nx^n,
 \eq
where
\bq c_n=\sum_m\frac{1}{m!}\b^m\tilde{c}_{m+n}.
 \eq
By choosing $\b$ appropriately we can consistently drop one of
the coupling parameters $c_n$.  A natural choice is to take $c_2=0$, which
requires \bq \sum_m\frac{1}{m!}\b^m\tilde{c}_{m+2}=0.
 \label{beqq} \eq
We will henceforth assume that $\b$ satisfies (\ref{beqq}).
It is  useful to exhibit explicitly the
three $x$-dependent functions which appear in the collective field
Lagrangian.  They are,
\brr W'(x) &=& \sqrt{N}c_1+\hf\frac{c_3}{\sqrt{N}}x^2
     +\frac{1}{6}\frac{c_4}{N}x^3+\cdots, \nonumber \\
     W'(x)^2 &=& Nc_1^2+c_1c_3x^2+\frac{1}{3}\frac{c_1c_4}{\sqrt{N}}x^3
     +\cdots, \nonumber \\
     W''(x) &=& \frac{c_3}{\sqrt{N}}x+\cdots.
 \label{drone} \err
For any finite $x$, all terms involving $c_n$ for $n\ge 4$ vanish as
$N\rightarrow\infty$.  Since this is the limit of interest in the
remaining part of this paper,
we can therefore, without
loss of generality, neglect all $c_n$ for $n\ge 4$.  Since $c_2$ has
independently been set to zero by shifting $x$, the most general
superpotential for our purposes is of the form
\bq W(x)=Nc_0+\sqrt{N}c_1x+\frac{1}{6}\frac{c_3}{\sqrt{N}}x^3.
 \label{potform}\eq
An important qualitative aspect of this superpotential depends on
the sign of the product $c_1c_3$.  Specifically, in the large
$N$ limit, the potential, $\hf W'(x)^2$, will be a parabola
which is concave up if $c_1c_3<0$ or concave down if $c_1c_3>0$.
The interesting physics, which we will discuss below, depends crucially
on the existence of a local maximum in this potential.
We will therefore take $c_1c_3<0$.

{\fl{\it 4.3 $N\rightarrow\infty$ Collective Field Theory}}

We now take the limit $N\rightarrow\infty, L\rightarrow\infty$.
As noted above, this limit can be taken in one of two ways.
We will discuss each of these possibilities in detail.
It is important to note that we may take the limit in either manner,
independently, within any given region of $x$.

a) ``Low Density" case:  The first possibility is that $N\rightarrow\infty,
L\rightarrow\infty$ but, over the range $x_1<x<x_1+l_1$, $N/l_1$
remains finite.  In this case the density of eigenvalues remains sparse.
Under this circumstance, within this region, the collective fields
(\ref{colldef}) contain only a finite number of independant modes.
The collective fields must then satisfy constraints which are simply the
definitions (\ref{colldef}). The collective field Lagrangian (\ref{lagcollpp})
applies to the physics in this region, but it must be understood that $\vphi$
and $\psi$ are constrained and this fact must be duly accounted for.  Because
of
the definitions (\ref{colldef}) and the fact that (\ref{lagcollpp})
is merely a rewriting of (\ref{lageig}), the natural way to
avoid this complication is to simply use (\ref{lageig}) to describe
the physics of the individual eigenvalues.  Any eigenvalue
behavior which is deduced using (\ref{lageig}) can then be cast in
collective field language by invoking (\ref{colldef}).

b) ``High Density" case:  The other possibility is that
$N\rightarrow\infty, L\rightarrow\infty$ such that, within a region
$x_2<x<x_2+l_2$, $N/l_2\rightarrow\infty$.  In this case the eigenvalues
become dense.  The collective fields become, modulo a subtlety to be
discussed below, unconstrained, ordinary
two dimensional fields.
In this limit the eigenvalue Lagrangian (\ref{lageig}) becomes
less useful.  This is because it is difficult to interpret
the sums over an infinite number of unspecified, dense eigenvalues.
The collective field Lagrangian (\ref{lagcollpp}) offers a more
useful description of the system.
However, some care must now
be taken in evaluating the last three terms of (\ref{lagcollpp}).
We would like to use equation (\ref{ppdef}) to perform the integrations
in these terms.  In the case of finitely seperated eigenvalues there
is not a problem, as discussed above.
However, for densely packed eigenvalues, these integrals diverge
and have to be regulated.  We propose a regulation
procedure which is implemented using properties of complex
integration.  In this way
sensible finite results can be
obtained, but there are important subtle ambiguities.  We proceed with
an analysis of this issue.

Consider an analytic function,
$\phi(z)$, where $z=x+iy$, and assume that
$\phi(z)\rightarrow 0$ as $z\rightarrow\infty$.
Over a contour which traverses the real axis, $x=(-\infty,+\infty)$
and then closes back in either the upper or lower half plane
we have,
\bq \oint dz\frac{\phi(z)}{z-a}=\int_{-\infty}^{+\infty}dx
    \frac{\phi(x)}{x-a}.
 \eq
This is because the contribution from the contour at infinity
vanishes.  Deform the contour around the pole
using using a semicircle of radius $\e$.

\ \vspace{2pc}

\centerline{\tenrm Figure~2. Contour of integration $C_+$}
We consider two
possibilities.  In the first, we choose a countour,
which we denote $C_+$, which
follows the real axis from $-\infty$
to the point $x=a-\e$, then follows a semicircle,
$\g_+$, around the pole in the upper half plane to the point
$x=a+\e$, follows the $x$ axis to $+\infty$, and then closes back
in the upper half plane. In the second case we consider the mirror
image contour, $C_-$ in the lower half plane.  The small semicircle
is then denoted $\g_-$.  The contour $C_+$ is depicted in Figure 2.
We can now apply the Cauchy-Riemann theorem.
Since the contribution at infinity vanishes, we see that
\bq \oint_{C_\pm}dz\frac{\phi(z)}{z-a}
    =\ppb dx\frac{\phi(x)}{x-a}
    +\lim_{\e\rightarrow 0}\int_{\g_\pm}dz\frac{\phi(z)}{z-a}.
 \label{cauch} \eq
It is easy to show, using polar coordinates, that
\bq \lim_{\e\rightarrow 0}\int_{\g_\pm}\frac{\phi(z)}{z-a}
    =\mp i\pi\phi(a).
 \eq
The left hand side of (\ref{cauch}) vanishes since the full contour
does not encompass any poles.  Thus,
\bq \ppb dx\frac{\phi(x)}{x-a}=\pm i\pi\phi(a).
 \label{equu} \eq
Note the sign ambiguity.  This is due to the ambiguity concerning which
of the two contours $C_\pm$ we can choose when performing the
integration.  Note also that the right hand side of (\ref{equu})
is imaginary.
This may appear peculiar but it is the only mathematically
consistent way to make finite sense out of this irregular integral.
The sign ambiguity has physical significance to
the collective field theory as we  demonstrate shortly.

We proceed to discuss the last three terms of (\ref{lagcollpp})
sequentially.  The first of these is evaluated as follows,
\brr \frac{1}{3}\pp dxdydz\frac{\vphi'(x)\vphi'(y)\vphi'(z)}
     {(x-y)(x-z)}
     &=& \frac{1}{3}\int dx\vphi'(x)
     (\pp dy\frac{\vphi'(y)}{x-y})(\pp dz\frac{\vphi'(z)}{x-z})
     \nonumber \\
     &=&\frac{1}{3}\int dx\vphi'(x)(\pm i\pi\vphi'(x))(\pm i\pi\vphi'(x))
     \nonumber \\
     &=& \pm\frac{\pi^2}{3}\int dx\vphi'(x)^3.
 \label{w1id} \err
In the second line of (\ref{w1id}) the two ambiguous signs are
independent so that the final result has an ambiguous sign.
This sign determines the signature of the
two-dimensional spacetime metric.
We next consider the term
\bq \pp dxdy\frac{\vphi'(x)\vphi'(y)}{x-y}W'(x),
 \label{odef}
\eq
which we denote by $\cal{O}$.
Using (\ref{equu}), we see that when $N\rightarrow\infty$
this expression becomes antihermitian.
The collective field Lagrangian must be Hermitian.
 Recall, however,  that the original definition of this
term is given by the left hand side of
(\ref{ee2}), which is real.  This term may then be
decomposed as follows,
\brr \sum_i\frac{\der w}{\der\l_i}\frac{\der W}{\der\l_i}
     &=& a\{\sum_i\frac{\der w}{\der\l_i}\frac{\der W}{\der\l_i}\}
         +(1-a)\{\sum_i\frac{\der w}{\der\l_i}
         \frac{\der W}{\der\l_i}\}^* \nonumber \\
     &=& a{\cal{O}}+(1-a){\cal{O}}^*.
 \label{alf} \err
where $a$ is an arbitrary real parameter. Therefore,
in the limit $N\rightarrow\infty$,
\bq \sum_i\frac{\der w}{\der\l_i}\frac{\der W}{\der\l_i}
    \rightarrow\pm (2 a-1)i\pi\int dx\vphi'(x)^2W'(x).
 \eq
We can then choose $a=1/2$ and this term vanishes.
That is, in the limit $N\rightarrow\infty$ we can take the
next to last term in (\ref{lagcollpp}) to be zero.
This is the unique consistent prescription which yields a Hermitian
collective field Lagrangian in this limit.
We now turn to the remaining term in the collective field
Lagrangian (\ref{lagcollpp}).
It is useful to  express the fermion fields in terms
of the real and imaginary parts,
\brr \psi &=& \frac{1}{\sqrt{2}}(\psi_1+i\psi_2) \nonumber \\
     \bar{\psi} &=& \frac{1}{\sqrt{2}}(\psi_1-i\psi_2).
 \err
The remaining term in the Lagrangian (\ref{lagcollpp}) then reads
\brr & & -\hf\pp dxdy\Biggl\{\frac{1}{(x-y)}\bar{\psi}(x)\psi'(y)
         -\frac{\vphi''(y)}{\vphi'(x)}\bar{\psi}(x)\psi(x)\Biggr\}
         \nonumber \\
     & & \hspace{.2in} =-\hf\pp dxdy\frac{1}{(x-y)}
         \Biggl\{\psi_1(x)\psi_1'(y)+\psi_2(x)\psi_2'(y) \nonumber \\
     & & \hspace{1.3in}
         +i\psi_1(x)\psi_2'(y)-i\psi_2(x)\psi_1'(y)
         -\frac{\vphi''(y)}{\vphi'(x)}\bar{\psi}(x)\psi(x)\Biggr\}.
 \label{lastex} \err
Using (\ref{equu}) we see that the last three terms in (\ref{lastex})
become antihermitian as $N\rightarrow\infty$.
Since the Lagrangian must be Hermitian, we treat this problem in
exact analogy with the previous discussion.  That is, we reexpress
(\ref{lastex}) in terms of the lefthand side of (\ref{ee3}).  This
real expression is then decomposed exactly as in (\ref{alf}), and
$a$ is chosen to be $1/2$ to cancel the antihermitian terms.
However, unlike the previous case the final result is nonvanishing.  We
find that, in the $N\rightarrow\infty$ limit, we can consistently
take the last term in (\ref{lagcollpp}) to be
\bq  \int dx\Biggl\{\pm\frac{i\pi}{2}\psi_1(x)\psi_1'(x)
     \pm\frac{i\pi}{2}\psi_2(x)\psi_2'(x)\Biggr\}.
 \eq
The two ambiguous signs in this expression
 are independant. As will be seen, the choice of these
signs determines the ``chiralities" of the two dimensional fermions.

We have now consistently interpreted the principle part terms in the
collective field Lagrangian in the ``dense" $N\rightarrow\infty,
L\rightarrow\infty$ limit.
Before exhibiting the full resultant Lagrangian we  note the
following  facts.  First,  from (\ref{drone}), we see that, as
$N\rightarrow\infty$,
\brr W''(x) &\rightarrow& 0 \nonumber \\
     W'(x)^2 &\rightarrow& \Lambda-\w^2x^2,
 \err
where $\Lambda=\hf N c_1^2$ and $\w^2=-c_1c_3$ are positive
constants.  It follows from the first of these expressions that
we can neglect the third term in (\ref{lagcollpp}).
Second,  recall that  the definition (\ref{colldef}) implies that
\bq \int \vphi'(x)dx=N.
 \label{constraint} \eq
This constraint continues to hold
in the large $N$ limit, despite the fact that $\vphi$ then has an
infinite number of independant
modes.  We cannot then arbitrarily vary
the Lagrangian to obtain the field equations.
The easiest way to handle this is to introduce constraint
(\ref{constraint}) into the Lagrangian by means of a Lagrange
multiplier, in which case field $\vphi$ becomes completely
unconstrained. The correct
procedure for doing this is described in detail in Appendix C.
The end result of this complicated procedure, however, is simply a
modification of the constant $\Lambda$. It turns out that
all subsequent results
are correct if we simply take $\Lambda=0$
and treat $\vphi$ as an unconstrained field.  This is the subtlety
concerning the continuous field $\vphi$ alluded to several times above.
The interested reader is referred to
Appendix C for a proof of this.

Combining (\ref{lagcollpp})
with all of the facts just
discussed, the effective $N\rightarrow\infty, L\rightarrow\infty$
high density collective field Lagrangian reads
\brr L &=& \int dx\Biggl\{\frac{\dot{\vphi}^2}{2\vphi'}
     \pm\frac{\pi^2}{6}\vphi^{'3}
     +\hf\w^2x^2\vphi' \nonumber \\
     & & \hspace{.3in}
     -\frac{i}{2\vphi'}(\psi_1\dot{\psi}_1+\psi_2\dot{\psi}_2)
     \pm\frac{i\pi}{2}\psi_1\psi_1'
     \pm\frac{i\pi}{2}\psi_2\psi_2' \nonumber \\
     & & \hspace{.3in}
     +\frac{i}{2}\frac{\dot{\vphi}}{\vphi^{'2}}
     (\psi_1\psi_1'+\psi_2\psi_2')\Biggr\}.
 \err
There are three notable features of this Lagrangian.
The first is that it is
neither translationally invariant, nor Lorentz invariant, nor
supersymmetric.
The second is that
the kinetic energy terms are not in canonical form, and the third
is that there are ambiguous signs.
The first issue, that the theory does not have the appropriate
symmetry for a realistic two-dimensional field theory will be resolved
in the next section.  For now we defer a discussion of this point.
The second issue is resolved in the next subsection when we canonicalize
the theory by redefining our fields and by redefining our spatial
coordinate.  We will now address the issue of the ambiguous signs.
There is physics in the choice of
each of these ambiguous signs.  The
first of them dictates the signature of the two-dimensional
spacetime metric,
which we desire to be Minkowskian. The appropriate choice for the first
ambiguous sign then turns out to be the minus sign.
The remaining two ambiguous signs dictate
chiralities for the respective fermions.
For reasons of supersymmetry to be discussed, we require that the two
fermion fields have opposite chirality.
We then choose the second ambiguous sign to be a minus and the
third to be a plus. To be clear, we rewrite the
collective field Lagrangian with these sign choices,
\brr L &=& \int dx\Biggl\{\frac{\dot{\vphi}^2}{2\vphi'}
     -\frac{\pi^2}{6}\vphi^{'3}
     +\hf\w^2x^2 \vphi' \nonumber \\
     & & \hspace{.3in}
     -\frac{i}{2\vphi'}(\psi_1\dot{\psi}_1+\psi_2\dot{\psi}_2)
     -\frac{i\pi}{2}\psi_1\psi_1'
     +\frac{i\pi}{2}\psi_2\psi_2' \nonumber \\
     & & \hspace{.3in}
     +\frac{i}{2}\frac{\dot{\vphi}}{\vphi^{'2}}
     (\psi_1\psi_1'+\psi_2\psi_2')\Biggr\}.
 \label{lagcollnoncan} \err
This expression is the $N\rightarrow\infty,
L\rightarrow\infty$ high density collective field Lagrangian which
is compatible with Minkowski spacetime and two dimensional supersymmetry.
Note that
the bosonic part of this Lagrangian is identical to the
bosonic collective field theory Lagrangian derived in \cite{bo}.
\pagebreak

{\fl{\it  4.4 Canonical High
Density Collective Field Theory}}

In order to identify the canonical fields of the theory
we have to shift the field $\vphi$ around a solution to
its equation of motion.  We then have to
perform a coordinate transformation.
In this subsection we will perform these operations and arrive at a
canonical Lagrangian for the high density collective field theory.
We begin by listing the equations of
motion derived from (\ref{lagcollnoncan}).  They are
\brr   & & \der_t(\frac{\dot{\vphi}}{\vphi'})
       -\hf\der_x(\frac{\dot{\vphi}^2}{\vphi^{'2}}+\pi^2\vphi^{'2}
       -\w^2x^2)   \nonumber \\
       & &\hspace{.4in}
       +\der_x\Biggl\{\frac{i}{2\vphi^{'2}}(\psi_1\dot{\psi}_1
       +\psi_2\dot{\psi}_2)-i\frac{\dot{\vphi}}{\vphi^{'3}}
       (\psi_1\psi_1'+\psi_2\psi_2')\Biggr\} \nonumber \\
       & &\hspace{.4in}
+\der_t\Biggl\{\frac{i}{2\vphi^{'2}}(\psi_1\psi_1'+\psi_2\psi_2')\Biggr\}
       = 0,  \nonumber\\
       &&
     \der_t(\frac{\psi_1}{\vphi'})-\der_x(\frac{\dot{\vphi}}{\vphi^{'2}}
       \psi_1-\pi\psi_1) = 0,  \nonumber\\
       & &
     \der_t(\frac{\psi_2}{\vphi'})-\der_x(\frac{\dot{\vphi}}{\vphi^{'2}}
       \psi_2+\pi\psi_2) = 0,
 \label{mott} \err
for the $\vphi$ field and for the $\psi_1, \psi_2$ fields, respectively.
We  focus on solutions $(\vphi,\psi_1,\psi_2)=(\tilde{\vphi}_0,
\tilde{\psi}_{10}, \tilde{\psi}_{20})$ which have the following
property, $\dot{\tilde{\vphi}}_0=\tilde{\psi}_{10}=\tilde{\psi}_{20}=0$.
(We denote classical solutions with both a tilde and a subscript $0$
for reasons to become clear below). That is,
we are interested in static, purely bosonic solutions.  The second two
equations
in (\ref{mott}) are then solved automatically and the first becomes \bq
\der_x(\pi^2\tilde{\vphi}_0^{'2}-\w^2x^2)=0.
 \eq
This  implies that
\bq \tilde{\vphi}_0'=\frac{1}{\pi}\sqrt{\w^2x^2-1/{\rm g}},
 \label{oldeq} \eq
where ${\rm g}$ is an arbitrary
integration constant.  However,
we are interested exclusively in the case ${\rm g}>0$, since this case
yields the interesting physics, as we will discuss.
It then follows that
$\tilde{\vphi}_0$ is only defined for $|x|\ge 1/(\w\sqrt{{\rm g}})$.
This is a very significant fact.  It turns out that we cannot
canonically define the dense collective field theory in the region
$|x|\le 1/(\w\sqrt{{\rm g}})$.
As we will see, there are other problems with this region as well.
Notably, the theory, properly expressed in terms of canonical
fields, possesses a space-dependent coupling parameter which actually
blows up at the boundaries of this region.  We will discuss this issue
at length below, but before proceeding we will say a few words about our
interpretation of this.  The high density collective field theory
is only valid in the region $|x|\ge 1/(\w\sqrt{g})$.  The infinite number
of eigenvalues
, defined over $x$, densely
populate only the ``exterior" region.  Within the region
$|x|\le 1/(\w\sqrt{g})$, exist only a finite number of eigenvalues.
Their behavior is described, not by the high
density collective field theory, but, more properly, by the eigenvalue
Lagrangian (\ref{lageigsusy}).  The actual mechanics of how the physics
in the different regions is patched together will be described later.
For the moment, we will continue to focus on the
high density theory which
is defined only in the regions $|x|\ge 1/(\w\sqrt{{\rm g}})$.

Equation (\ref{oldeq}) can be integrated.  Doing this we find
the most general purely bosonic, static solution to the equations of
motion derived from the high density
collective field theory.  The result is
\brr \tilde{\vphi}_0(x)=
     \left\{\begin{array}{cc}
            a_-+\frac{x}{2\pi}\sqrt{\w^2x^2-1/{\rm g}}
            +\frac{1}{2\pi\w {\rm g}}
            \ln(-\sqrt{\w}x+\sqrt{\w x^2-1/{\rm g}\w})
            & \mbox{$;  x\le \frac{-1}{\w\sqrt{{\rm g}}}$} \\
            a_++\frac{x}{2\pi}\sqrt{\w^2x^2-1/{\rm g}}
            -\frac{1}{2\pi\w {\rm g}}
            \ln(+\sqrt{\w}x+\sqrt{\w x^2-1/{\rm g}\w})
            & \mbox{$;  x\ge \frac{+1}{\w\sqrt{{\rm g}}}$}
      \end{array}\right. .
 \label{phix} \err
The parameters $a_+$ and $a_-$ are independant arbitrary integration
constants.
We now take the solution $\tilde{\vphi}_0$ as a background and define
a new field, $\z$, as the fluctuation around this background,
\bq \vphi=\tilde{\vphi}_0(x)+\frac{1}{\sqrt{\pi}}\z.
 \eq
Expressed in terms of the shifted field, $\z$, the Lagrangian
(\ref{lagcollnoncan})  reads
\brr L &=& \int dx\Biggl\{ \frac{ \frac{1}{\sqrt{\pi}}\dot{\z}^2              }
                { 2(\tilde{\vphi}'_0(x)+\frac{1}{\sqrt{\pi}}\z')}
       -\frac{\sqrt{\pi}}{6} \z^{'3}
       -\frac{\pi}{2}\tilde{\vphi}_0'(x)\z^{'2} \nonumber \\
       & & \hspace{.3in}
       -\frac{i                                            }
             { 2(\tilde{\vphi}'_0(x)+\frac{1}{\sqrt{\pi}}\z') }
       (\psi_1\dot{\psi}_1+\psi_2\dot{\psi}_2)
       -\frac{i\pi}{2}\psi_1\psi_1'
       +\frac{i\pi}{2}\psi_2\psi_2' \nonumber \\
       & & \hspace{.3in}
       +\frac{i}{2}\frac{ \frac{1}{\sqrt{\pi}}\dot{\z} }
                   { (\tilde{\vphi}_0'(x)+\frac{1}{\sqrt{\pi}}\z')^2}
       (\psi_1\psi_1'+\psi_2\psi_2')\Biggr\}
       +\frac{\pi^2}{3}\int dx\tilde{\vphi}_0'(x)^3
 \err
It is now possible to perform a coordinate transformation in
order to render both the bosonic and fermionic kinetic energies
canonical.  The appropriate choice is to define a spatial
coordinate $\tau$ by
\brr \tau'(x) &=& \frac{1}{\pi}
     (\tilde{\vphi}_0'(x))^{-1} \nonumber \\
     &=& \frac{1}{\sqrt{\w^2x^2-1/g}}.
 \err
Integrating this, we find
\brr \tau(x)=
     \left\{\begin{array}{cc}
            (\tau_0-\frac{\s}{2})-\frac{1}{\w}
            \ln(\sqrt{g\w^2x^2}+\sqrt{g\w^2x^2-1})
            & \mbox{$;  x\le \frac{-1}{\w\sqrt{g}}$} \\
            (\tau_0+\frac{\s}{2})+\frac{1}{\w}
            \ln(\sqrt{g\w^2x^2}+\sqrt{g\w^2x^2-1})
            & \mbox{$;  x\ge \frac{+1}{\w\sqrt{g}} $}
      \end{array}\right. ,
 \label{tref} \err
where $\tau_0$ and $\s$ are independant integration constants.
Although the dense collective field theory is not defined in the
region $|x|<\frac{1}{\w\sqrt{g}}$, we can, and will, continue the
definition of $\tau$ into this region.  We require that
$\tau(x)$ and $\tau'(x)$ match at the boundary.  The following
is then a suitable choice,
\brr \tau(x)=
     \left\{\begin{array}{cl}
            (\tau_0-\frac{\s}{2})
            -\frac{1}{\w}
            \ln(\sqrt{g\w^2x^2}+\sqrt{g\w^2x^2-1})
            & \mbox{$;  x\le \frac{-1}{\w\sqrt{g}}$} \\
            \tau_0
            +\frac{\s}{\pi}\sin^{-1}(x\w\sqrt{g})
            & \mbox{$;  \frac{-1}{\w\sqrt{g}}<x<\frac{+1}{\w\sqrt{g}}$}\\
            (\tau_0+\frac{\s}{2})
            +\frac{1}{\w}
            \ln(\sqrt{g\w^2x^2}+\sqrt{g\w^2x^2-1})
            & \mbox{$;  x\ge \frac{+1}{\w\sqrt{q}} $}
      \end{array}\right. .
 \label{diffeo} \err
The inverse of this transformation is given by
\brr x(\tau)=
     \left\{\begin{array}{cl}
            \frac{-1}{\w\sqrt{g}}
            \cosh\{\w(\tau-\tau_0+\s/2)\}
            & \mbox{$; \tau\le
            (\tau_0-\frac{\s}{2})
            $} \\
            \frac{1}{\w\sqrt{g}}\sin\{\frac{\pi}{\s}
            (\tau-\tau_0)\}
            & \mbox{ $;
            (\tau_0-\frac{\s}{2})
            <\tau<
            (\tau_0+\frac{\s}{2})
            $}\\
            \frac{+1}{\w\sqrt{g}}
            \cosh\{\w(\tau-\tau_0-\s/2)\}
            & \mbox{ $; \tau\ge
            (\tau_0+\frac{\s}{2})
            $ }
      \end{array}\right. .
 \label{xoft}\err
This transformation is depicted in figure 3.

\ \vspace{2pc}

\centerline{\tenrm Figure~3. The $x-\tau$ transformation.}
It is easily seen that $\tau_0$ is the position,
in $\tau$ space, of the center of the low density region and that
$\s$ is the width of this region.
We may now express the background solution in terms
of $\tau$. To avoid confusion,
we define $\vphi_0(\tau)=\tilde{\vphi}_0(x(\tau))$.
This explains the use of the tilde.
Using (\ref{phix}) and
(\ref{diffeo}), this is
\brr \vphi_0(\tau)=
     \left\{\begin{array}{cc}
            a_-+\frac{1}{2\pi g}(\tau-\tau_0+\frac{\s}{2})
            +\frac{1}{4\pi\w g}
            \sinh\{2\w(\tau-\tau_0+\frac{\s}{2})\}
            & \mbox{$;  \tau\le
            (\tau_0-\frac{\s}{2})
            $} \\
            a_++\frac{1}{2\pi g}(\tau-\tau_0-\frac{\s}{2})
            +\frac{1}{4\pi\w g}
            \sinh\{2\w(\tau-\tau_0-\frac{\s}{2})\}
            & \mbox{$; \tau\ge
            (\tau_0+\frac{\s}{2})
            $}
      \end{array}\right. .
 \label{rudolph} \err

In the region $|x|\ge 1/(\w\sqrt{g})$,
it is useful to define a function
\brr \tilde{\gbig}(x) &=& \pi^{-3/2} (\tilde{\vphi}_0'(x))^{-2}
     \nonumber\\
      &=& \frac{\sqrt{\pi}}{\w^2x^2-\frac{1}{g}}.
 \err
In terms of $\tau$, we then have
$\gbig(\tau)=\tilde{\gbig}(x(\tau))$, which reads
\brr \gbig(\tau)=
     \left\{\begin{array}{ll}
            \gbig_-(\tau) & \mbox{$;  \tau\le(\tau_0-\s/2)$ } \\
            \gbig_+(\tau) & \mbox{$;  \tau\ge(\tau_0+\s/2)$ }
      \end{array}\right.,
 \label{ggg} \err
where
\bq \gbig_\pm(\tau)=
     4\sqrt{\pi}g\frac{\frac{1}{\k} e^{\mp 2\w(\tau-\tau_0)}}
     {(1-\frac{1}{\k}e^{\mp 2\w(\tau-\tau_0)})^2},
 \label{gdef}\eq
and $\k$ is a dimensionless constant,
\bq \k=\exp{(\w\s)},
 \label{kdef} \eq
which relates the width, $\s$, of the low density region
in $\tau$ space to the natural length scale in the matrix model,
$1/\w$.
As we will see momentarily, function (\ref{gdef})
is the coupling parameter
in the high density collective field theory, expressed, canonically, in
$\tau$ space.  It is a space-dependent coupling, and is plotted
in figure 4.

\centerline{\tenrm Figure~4. Space dependent coupling parameter
            $\gbig_\pm$.}

Before exhibiting the collective field Lagrangian in $\tau$ space,
we will first discuss one small issue.  That is, in $\tau$ space,
the fermionic kinetic energy has the correct normalization
only if we trivially scale the fields $\psi_1$ and $\psi_2$.
Toward this end, we define
\brr \psi_+ &=& \frac{2^{1/4}}{\sqrt{\pi}}\psi_1 \nonumber \\
     \psi_- &=& \frac{2^{1/4}}{\sqrt{\pi}}\psi_2.
 \label{psidef} \err
The ``$\pm$" is a useful notation in two dimensions.
As we will see, this designation relates
to the Lorentz structure of these fields. Now, using the coordinate
transformation (\ref{diffeo}), and the definitions (\ref{gdef}),
(\ref{psidef}),
we can write the high density collective field Lagangian
as follows,
\brr L &=& \int d\tau\Biggl\{
       \hf(\dot{\z}^2-\z^{'2})
       -\frac{i}{\sqrt{2}}(\psi_+\dot{\psi}_+-\psi_+\psi_+')
       -\frac{i}{\sqrt{2}}(\psi_-\dot{\psi}_-+\psi_-\psi_-')
       \nonumber \\
       & & -\hf\frac{\gbig(\tau)\dot{\z}^2\z'}{1+\gbig(\tau)\z'}
       -\frac{1}{6}\gbig(\tau)\z^{'3} \nonumber \\
       & &
       +\frac{i}{\sqrt{2}}\frac{\gbig(\tau)\z'}{1+\gbig(\tau)\z'}
       (\psi_+\dot{\psi}_++\psi_-\dot{\psi}_-) \nonumber \\
       & & +\frac{i}{\sqrt{2}}\frac{\gbig(\tau)\dot{\z}}
       {(1+\gbig(\tau)\z')^2}(\psi_+\psi_+'+\psi_-\psi_-')\Biggr\}
       +\frac{1}{3}\int d\tau\frac{1}{\gbig(\tau)^2},
 \label{lagcollfinal} \err
where now the prime means $\der/\der\tau$.  This  is the
$N\rightarrow\infty, L\rightarrow\infty$ high density collective
field Lagrangian with canonically normalized kinetic energy terms.
The bosonic terms of this Lagrangian are
identical to the canonical bosonic collective field theory.
We reiterate that (\ref{lagcollfinal}) is only valid in the
high density regions $\tau\le(\tau_0-\s/2)$ and
$\tau\ge(\tau_0+\s/2)$.  In the region
$(\tau_0-\s/2)<\tau<(\tau_0+\s/2)$ there are only a finite number
of eigenvalues, whose dynamics is best described by Lagrangian
(\ref{lageigsusy}).

\renewcommand{\theequation}{5.\arabic{equation}}
\setcounter{equation}{0}
{\fl{\bf 5.  The Supersymmetric Effective Theory}}

The Lagrangian (\ref{lagcollfinal}) has kinetic energy terms for both
the bosonic field, $\z$ and for the fermionic fields,
$\psi_\pm$, which are canonically normalized for a flat
two-dimensional spacetime.  The interaction terms, however,
involve an explicit spatially-dependent coupling, $\gbig(\tau)$,
which violates Poincare invariance.  We interpret $\gbig(\tau)$ to be
the vacuum expectation value (VEV) of a function of an additional
``heavy"
field, which we denote by $\a$. Furthermore, we
infer the existence of an effective
theory involving $\a$, as well as $\z$, $\psi_+$ and $\psi_-$, which
reproduces (\ref{lagcollfinal}) when $\a$ is replaced by its
$\tau$-dependent VEV.
Additionally, we postulate that the effective theory possesses a
two-dimensional supersymmetry.  It follows that, in addition to
$\a$, we must introduce its fermionic superpartners, $\chi_+$ and
$\chi_-$ which, of course, have vanishing VEV's.
The field $\a$ and its superpartners $\chi_+$ and $\chi_-$ are
assumed to be heavy. We do not consider their fluctuations
but rather treat them as frozen in their VEV's.
Although we have a ready interpretation of the light
fields $\zeta, \psi_+$, and $\psi_-$ as being comprised
of modes related to the singlet sector of the underlying
matrix model, we do not attempt a similar
interpretation of the heavy fields. A precise explanation for
treating these fields as suggested is beyond the scope of this
paper. We can only offer at this point a motivation from two dimensional
string theory. Two dimensional string theory
contains, in its associated effective low-energy Lagrangian,
a massive dilaton multiplet which is
frozen at its VEV as a consequence of gauge symmetry.
Furthermore,
although the underlying matrix theory is supersymmetric,
it is not obvious
that our effective two dimensional
theory must also be supersymmetric.
Nevertheless, we proceed in this section to
demonstrate that, indeed, there exists an effective
two-dimensional theory which
is Poincare invariant and supersymmetric,
which involves both heavy fields
and light fields and which, when the heavy fields are replaced by their
VEV's, reproduces the collective field Lagrangian (\ref{lagcollfinal}).

{\fl {\it 5.1 Two Dimensional (p,q) Supersymmetry}}

It is well known that, because the two-dimensional Lorentz group
is abelian, the possible $d=2$ supersymmeties have a rich structure
\cite{hullwitten,bebmg}. Specifically,
there exist supersymmetries with any number of left-chiral fermionic
generators, $Q_{A-}, A=1,...,p$, and any independent number, $q$, of
right-chiral fermionic generators, $Q_{I+}, I=1,..., q$.
There are thus an infinite number of
two-dimensional supersymmetries enumerated by the respective
numbers of left- and right-chiral fermionic
generators, $p$ and $q$.  These
are called $(p,q)$ supersymmetries.  The generators must satisfy
the following algebra,
\brr \{Q_{A-},Q_{B-}\} &=& -2i\d_{AB}\der_-  \nonumber \\
     \{Q_{I+},Q_{J+}\} &=& -2i\d_{IJ}\der_+  \nonumber \\
     \{Q_{A-},Q_{I+}\} &=& 0,
 \err
where indices $A,B,...$ run from 1 to $p$ and the indices $I,J,...$
run from 1 to $q$.
Which of these, if any, is the appropriate supersymmetry of our
theory ?

We may quickly narrow
the range of possibilities by the following observations.
The collective field Lagrangian, (\ref{lagcollfinal}), describes all light
fields in the effective theory.  Since the collective field Lagrangian has only
one bosonic, $\z$, and two fermionic, $\psi_\pm$, degrees of freedom, and since
the light fields and heavy fields cannot belong to
the same supersymmetric multiplet, it follows that
the number of fermions in the fundamental matter
multiplet of the relevant supersymmetry is at most two. Since
the number of fermions in the fundamental matter multiplet is the same
as the
number of supersymmetry generators, it follows
that $p+q\le 2$.  There are, therefore, only five possibilities,
$(1,0), (0,1), (2,0), (0,2)$, or $(1,1)$ supersymmetry.
We will now examine each of these possibilities in turn.

a) (1,0) and (0,1) supersymmetry:
There is only one matter multiplet for either
(1,0) or (0,1) supersymmetry.  It contains one boson and one fermion.
The light sector of the effective theory would need two such
multiplets to accomodate the required number of fermions.  The theory
would then have two bosons as well as two fermions.  Since we require
that there be only one light boson, it is impossible to
properly describe the necessary degrees of freedom using (1,0) or
(0,1) supersymmetry.  This case is thus ruled out.

b) (2,0) and (0,2) supersymmetry:
A (0,2) superspace has coordinates
$z^M=(x^\pm,\t^+,\bar{\t}^+)$ where $\t^+$ and $\bar{\t}^+$ are
complex conjugates.  The covariant superspace derivatives are given by
\brr D_+ &=& \frac{\der}{\der\t^+}+i\bar{\t}^+\der_+
         \nonumber \\
     \bar{D}_+ &=& \frac{\der}{\der\bar{\t}}+i\t^+\der_+
 \err
Note that $(\der/\der\t)^*=-(\der/\der\bar{\t})$, and
$\bar{D}=-D^*$. An irreducible ``chiral" superfield, $\Phi$ is
obtained by imposing the differential constraint $\bar{D}_+\Phi=0$.
In component fields such a superfield reads
\bq \Phi=\phi+i\sqrt{2}\t^+\psi_++i\t^+\bar{\t}^+\der_+\phi,
 \label{sup1} \eq
where $\phi$ and $\psi$ are complex.  The complex conjugate,
$\bar\Phi$, satisfies $D_+\bar\Phi=0$ and is called
``antichiral".  It is given by
\bq \bar\Phi=\phi^*+i\sqrt{2}\bar{\t}^+\bar{\psi}_+
    -i\t^+\bar{\t}^+\der_+\phi^*.
 \label{sup2} \eq
(We recall that $(\t\psi)^*=\bar{\psi}\bar{\t}=-\bar{\t}\bar{\psi}$.)
The (0,2) transformation law, on the component fields, is
\brr \d\phi &=& i\n^+\psi_+  \nonumber \\
     \d\psi_+ &=& \bar{\n}^+\der_+\phi \nonumber \\
     \d\phi^* &=& i\bar{\n}^+\bar{\psi}_+  \nonumber \\
     \d\bar{\psi}_+ &=& \n^+\der_+\phi^*.
 \err
It can be shown that these are the only irreducible
representations of (0,2) supersymmetry.
We exhibit the above details in order to allow the following possibility.
First of all, we note that the irreducible representations of (0,2)
supersymmetry each have two bosonic and two fermionic degrees of freedom.
By our reasoning above, we should rule out this supersymmetry from
consideration immediately since the effective theory is desired to
have only one light boson.  However, it is tempting to try to accomodate
the new boson $\a$ along with the light fields $\z$, $\psi_+$ and
$\psi_-$ in a single supersymmetric multiplet.
We would then have to find some alternative explanation
for neglecting the fluctuations of the extra boson,
but this would avoid the otherwise necessary addition of extra
heavy fermions.
The (0,2) multiplet is particularly suited to this idea because its
fundamental representation has exactly two bosons and two fermions.
It turns out, however, after an extensive analysis of this
possibility, that this supersymmetry can be ruled out even if we allow
this last idea. This is because it is
imposible to construct an interaction Lagrangian using the superfields
(\ref{sup1}) and (\ref{sup2}) which has the appropriate derivative structure
indicated in (\ref{lagcollfinal}).  This problem is related to the existence of
the derivative in the highest components of $\Phi$ and $\bar\Phi$.  The same
reasoning applies to (2,0) supersymmetry. We can thus rule out
(2,0) and (0,2) as candidate supersymmetries of the effective theory.

c) (1,1) supersymmetry:
The sole remaining possibility is (1,1) supersymmetry.
As we will show, it is indeed possible to construct an effective theory
with this supersymmetry and with the desired relationship
to the collective field theory derived in the last section.
We proceed to describe (1,1) supersymmetry in some detail.  We will
then derive the effective theory, discuss its equations of motion,
solve these equations of motion, and show that when the heavy fields
are replaced by their VEV's that the high density
collective field Lagrangian (\ref{lagcollfinal}) is recovered.

{\fl{\it 5.2 (1,1) Supersymmetry}}

A (1,1) superspace has coordinates $z^M=(x^\pm,\t^+,\t^-)$.
The supersymmetry generators are given by
\brr Q_+ &=& \frac{\der}{\der\t^+}-i\t^+\der_+ \nonumber \\
     Q_- &=& \frac{\der}{\der\t^-}-i\t^-\der_-.
 \err
and satisfy the algebra,
\brr \{Q_\pm,Q_\pm\} &=& -2i\der_\pm \nonumber \\
     \{Q_+,Q_-\} &=& 0.
 \err
The covariant superspace derivatives are
\brr D_+ &=& \frac{\der}{\der\t^+}+i\t^+\der_+ \nonumber \\
     D_- &=& \frac{\der}{\der\t^-}+i\t^-\der_-.
 \err
The fundamental irreducible
representation is a real superfield,
$\Phi_1=\bar{\Phi}_1$, which, in component fields, is given by
\bq \Phi_1=\z+i\t^+\psi_++i\t^-\psi_-+i\t^+\t^-Z,
 \eq
where $\z$ and $Z$ are real and commuting and
$\psi_+$ and $\psi_-$ are real anticommuting spinors.
The (1,1) transformation law for the component fields is
\brr \d\z &=& i\n^+\psi_++i\n^-\psi_- \nonumber \\
     \d\psi_+ &=& \n^+\der_+\z+\n^-Z \nonumber \\
     \d\psi_- &=& \n^-\der_-\z-\n^+Z \nonumber \\
     \d Z &=& i\n^-\der_-\psi_+-i\n^+\der_+\psi_-.
 \err
Depending on the dynamics of the theory, the highest component
of this multiplet can be either  a nonphysical auxiliary degree of freedom
or  a physical propagating field.  The light sector is required
to have one physical boson and two physical fermions.  This can be accomodated
by the superfield $\Phi_1$ provided field $Z$ is auxiliary.
We therefore choose the Lagrangian so that this is the case.
We also require the existence of a
massive sector which includes the bosonic field, $\a$.  We must
then introduce a second, ``heavy" superfield, $\Phi_2$, given,
in components, as follows,
\bq \Phi_2=\a+i\t^+\chi_++i\t^-\chi_-+i\t^+\t^-A.
 \eq
As in $\Phi_1$, $\a$ is a real, physical boson, $\chi_+$ and
$\chi_-$ are real, physical fermions, and $A$ is an additional
boson whose status as auxiliary or physical  depends
on the form of the effective Lagrangian.  The
(1,1) transformation law on the component fields of $\Phi_2$ reads
\brr \d\a &=& i\n^+\chi_++i\n^-\chi_- \nonumber \\
     \d\chi_+ &=& \n^+\der_+\a+\n^-A \nonumber \\
     \d\chi_- &=& \n^-\der_-\a-\n^+A \nonumber \\
     \d A &=& i\n^-\der_-\chi_+-i\n^+\der_+\chi_-.
 \err
We would like to use the two superfields, $\Phi_1$ and $\Phi_2$, to
construct a (1,1) supersymmetric Lagrangian that  reproduces the high density
collective field Lagrangian, (\ref{lagcollfinal}), when the heavy fields are
replaced by their VEV's.  The discussion so far in this section has
demonstrated
conclusively that this is the minimal prescription which could concievably
satisfy this criterion.

{\fl {\it 5.3  The (1,1) Supersymmetric Effective Theory}}

We proceed to construct the effective theory using the two
(1,1) superfields introduced above.
For convenience, we list these superfields again,
\brr \Phi_1 &=& \z+i\t^+\psi_++i\t^-\psi_-+i\t^+\t^-Z  \nonumber \\
     \Phi_2 &=& \a+i\t^+\chi_++i\t^-\chi_-+i\t^+\t^-A.
 \err
Using these two superfields and the differential operators,
$D_+, D_-, \der_+$, and $\der_-$, we  build the effective
theory piecemeal, order by order in the coupling, $\gbig(\tau)$.  We
 begin with the free  part of the collective
field Lagrangian, (\ref{lagcollfinal}), and find the relevant
supersymmetric expression involving $\Phi_1$ and $\Phi_2$
which reproduces it when the equations of motion are
used.  We  then include the part of (\ref{lagcollfinal})
which is linear in $\gbig(\tau)$ and modify our construction
appropriately.  Proceeding in this manner, we  eventually
discover the entire supersymmetric effective theory.  We  end
this section by exhibiting the complete effective theory, listing
its equations of motion, and showing that when the equations of
motion are used, the collective field Lagrangian is recovered.

{\fl a) {\it $0^{th}$ order: }}

The free (noninteracting) part of the collective field Lagrangian,
(\ref{lagcollfinal}), involving the fields $\z, \psi_+$, and
$\psi_-$ is given by
\bq \L_{01}=\hf(\dot{\z}^2-\z^{'2})-i\psi_+\der_-\psi_+
         -i\psi_-\der_+\psi_-.
 \label{oops} \eq
This can be written manifestly supersymmetrically using $\Phi_1$.
The appropriate super-Lagrangian is
\brr \L_{01}^{(eff)} &=& \int d\t^+d\t^-D_+\Phi_1D_-\Phi_1 \nonumber \\
        &=& \hf(\dot{\z}^2-\z^{'2})-i\psi_+\der_-\psi_+
         -i\psi_-\der_+\psi_-+Z^2.
 \label{lago1} \err
Field $Z$ has no dynamics.  It is therefore auxiliary and can be
eliminated using its equation of motion, which reads $Z=0$.
Eliminating $Z$, (\ref{lago1}) becomes
\bq \L_{01}^{(eff)}=\hf(\dot{\z}^2-\z^{'2})-i\psi_+\der_-\psi_+
         -i\psi_-\der_+\psi_-,
 \eq
which is precisely the free collective field Lagrangian for $\z, \psi_+$,
and $\psi_-$ given in (\ref{oops}).

We also
need to introduce kinetic energy for the heavy multiplet, $\Phi_2$.
Additionally, we need to introduce a mass term or some suitable
alternative interaction for $\Phi_2$.  Since the last
term in (\ref{lagcollfinal}) involves only $\gbig(\tau)$, we presume
that it is the vestige of the pure $\Phi_2$ kinetic energy
Lagrangian which
must become a function of $\tau$ only
when the equations of motion are used.  Using (\ref{gdef}), the
last term in (\ref{lagcollfinal}),
$(3\gbig(\tau)^2)^{-1}$, can
be written as follows
\bq \L_{02}=\frac{\k^2}{48\pi {\rm g}^2}(e^{\w|\tau-\tau_0|}
-\frac{1}{\k}e^{-\w|\tau-\tau_0|})^4.
 \eq
We pose the important hypotheses that the equations of motion
admit the following solution,
\brr  <\a>       &=& e^{-\w|\tau-\tau_0|} \nonumber \\
      <\chi_\pm> &=& 0 \nonumber\\
      <A>        &=& 0.
\label{moon} \err
This amounts to the requirement that
\bq <\L_{\a}^{(eff)}>=\frac{\k^2}{48\pi {\rm g}^2}
    (<\a^{-1}>-\frac{1}{\k}<\a>)^4.
 \label{late} \eq
It is expedient to first construct a relevant Lagrangian involving
only $\a$, which we call $\L_\a^{(eff)}$, and then to
supersymmetrize the result to include the full multiplet, $\Phi_2$.
We must first decide what sort of structure we expect
$\L_\a^{(eff)}$ to have.  An obvious guess would be
\bq F_1(\a)\der_+\a\der_-\a-\w^2F_2(\a),
 \label{guess} \eq
where $F_1(\a)$ and $F_2(\a)$ are polynomials in $\a$.  It turns out
that there do exist $F_1$ and $F_2$ such that
(\ref{guess}) is both compatible with the exponential solution
(\ref{moon}) and with the desired property that $<\L_\a^{(eff)}>=
\L_{02}$. There is a convoluted impediment to the supersymmetrization
of this choice, however.  This is related to the fact that the
supersymmetrization of $\a^n$, which is $\int d\t^+d\t^-\Phi_2^m$,
generates interactions of the sort $\a^{m-1}A$.  The equation of motion
for $A$ then involves powers of $\a$, that is $<A>\ne 0$.  Under this
circumstance it is impossible to construct a Lagrangian which
reproduces the necessary higher-order interactions between
$\Phi_1$ and $\Phi_2$.  An interesting resolution to this problem
is the following.  We take
\bq \L_\a^{(eff)}= F_1(\a)\der_+\a\der_-\a
    -\frac{1}{\w^2}F_2(\a)(\der_+\der_-\a)^2
 \label{widdle}\eq
That is, we consider higher derivative interactions.  As we will see,
this generalizes supersymmetrically in such a way
that $<A>=0$.  How then do we
determine the functions $F_1(\a)$ and $F_2(\a)$?  There are two
criteria for this which, together, uniquely specify these
functions. First, the equation of motion derived from
(\ref{widdle}) need allow (\ref{moon}) as a solution, and, secondly,
using this solution, we must have $<\L_\a^{(eff)}>=\L_{02}$.
To resolve the first issue, we compute the equation of motion
using (\ref{widdle}).  This reads,
\brr & & 2F_1(\a)\der_+\der_-\a \nonumber \\
     & & \hspace{.4in}
         +F_1'(\a)\der_+\a\der_-\a \nonumber \\
     & & \hspace{.4in}
         +\frac{2}{\w^2}F_2(\a)\der_+^2\der_-^2\a \nonumber \\
     & & \hspace{.4in}
         +\frac{1}{\w^2}F_2'(\a)\Biggl\{2\der_+\a\der_+\der_-^2\a
         +(\der_+\der_-\a)^2
         +2\der_+(\der_-\a\der_+\der_-\a)\Biggr\} \nonumber \\
     & & \hspace{.4in}
         +\frac{2}{\w^2}F_2''(\a)\der_+\a\der_-\a\der_+\der_-\a=0.
 \label{mot1}\err
If $\a=\exp{(-\w|\tau-\tau_0|)}$, then this equation becomes
\bq 4\a F_1(\a)+2\a^2F_1'(\a)-2\a F_2(\a)-7\a^2F_2'(\a)
    -2\a^3F_2''(\a)=0.
 \label{mot2}\eq
If we express $F_1$ and $F_2$ as follows
\brr F_1(\a) &=& \sum_n a_n\a^n \nonumber \\
     F_2(\a) &=& \sum_n b_n\a^n,
 \err
then (\ref{mot2}) requires that
\bq b_n=\frac{2}{1+2n}a_m.
 \label{bin} \eq
To resolve the second issue, we can use (\ref{bin}) to rewrite
equation (\ref{widdle}) as
\bq \L_\a^{(eff)}=\sum_n a_n\a^n\Biggl\{\der_+\a\der_-\a
    -\frac{2}{1+2n}\cdot\frac{1}{\w^2}(\der_+\der_-\a)^2\Biggr\}.
 \eq
Substituting (\ref{moon}), we find that
\brr <\L_\a^{(eff)}> &=& -\w^2\sum_na_n\frac{1+n}{1+2n}
     <\a>^n \nonumber \\
     &=& -\w^2\sum_nc_n<\a>^n.
 \label{gnome} \err
Combining (\ref{bin}) and (\ref{gnome}), we find
\brr a_n &=& \frac{1+2n}{1+n}c_{n+2} \nonumber \\
     b_n &=& \frac{2}{1+n}c_{n+2}.
 \label{arx} \err
The coeffiecients $c_n$ are easily determined
from (\ref{late}).  The coefficients $a_n$ and $b_n$ are then computed
with (\ref{arx}).  This then determines both $F_1(\a)$ and
$F_2(\a)$.  The result is that
\brr F_1(\a) &=&-\frac{\k^2}{48\pi {\rm g}^2\w^2}
     (\frac{11}{5}\a^{-6}-\frac{28}{3}\frac{1}{\k}\a^{-4}
     +18\frac{1}{\k^2}\a^{-2}-4\frac{1}{\k^3}
     +\frac{5}{3}\frac{1}{\k^4}\a^2) \nonumber \\
     F_2(\a) &=& -\frac{\k^2}{48\pi {\rm g}^2\w^2}
     (-\frac{2}{5}\a^{-6}+\frac{8}{3}\frac{1}{\k}\a^{-4}
     -12\frac{1}{\k^2}\a^{-2}-8\frac{1}{\k^3}
     +\frac{2}{3}\frac{1}{\k^4}\a^2).
 \err
These functions look rather peculiar.  This is a consequence of the
fact that equation (\ref{widdle}) is not a unique
prescription for determining the pure $\a$  Lagrangian.
We could, for instance, have included interactions
of the form $\w^{-P}(\der_+\der_-\a)^P$
where $P$ is a completely arbitrary
exponent.  The corresponding $\a$  Lagrangian would
then be different. Our construction is, however, the simplest
example which has the appropriate relationship to the collective
field theory.  The supersymmetrization of (\ref{widdle}) is given
by the following super-Lagrangian,
\bq \L_{02}^{(eff)}=\int d\t^+d\t^-\Biggl\{
    F_1(\Phi_2)D_+\Phi_2D_-\Phi_2
    -\frac{1}{\w^2}F_2(\Phi_2)\der_-D_+\Phi_2\der_+D_-\Phi_2\Biggr\}.
 \label{lago2} \eq
In components this reads
\brr \L_{\a}^{(eff)}
    &=& F_1(\a)\der_+\a\der_-\a-\frac{1}{\w^2}F_2(\a)(\der_+\der_-\a)^2
    \nonumber \\
    & & +F_1(\a)A^2-\frac{1}{\w^2}F_2(\a)\der_+A\der_-A \nonumber \\
    & & -iF_1(\a)\chi_+\der_-\chi_+
        -\frac{i}{\w^2}F_2(\a)\der_-\chi_+\der_+\der_-\chi_+
        \nonumber \\
    & & -iF_1(\a)\chi_-\der_+\chi_-
        -\frac{i}{\w^2}F_2(\a)\der_+\chi_-\der_-\der_+\chi_-
 \err
Clearly, the equations of motion admit solution (\ref{moon}).
Also, $<\L_{\a}^{(eff)}>=\L_{02}$
by construction.

{\fl{\it b) $1^{st}$ order:}}\\
The collective field Lagrangian (\ref{lagcollfinal})
can be expanded in powers of $\gbig(\tau)$,
\bq \L=(\L_{01}+\L_{02})+\L_1+\L_2+\cdots.
 \eq
We have supersymmetrized $\L_{01}$ and $\L_{02}$ above.
The term linear in $\gbig(\tau)$ reads
\brr \L_1 &=&\gbig(\tau)\Biggl\{-\frac{1}{6}(\z^{'3}+3\dot{\z}^2\z') \nonumber
\\
     & & \hspace{.4in}
     +\frac{i}{\sqrt{2}}\z'(\psi_+\dot{\psi}_++\psi_-\dot{\psi}_-)
     \nonumber \\
     & & \hspace{.4in}
     +\frac{i}{\sqrt{2}}\dot{\z}(\psi_+\psi_+'+\psi_-\psi_-')\Biggr\}.
 \label{laglin} \err
This may be extended supersymmetrically as follows.  First of all
we note that
\brr \der_{(+}\Phi_1\der_{-)}\Phi_2| &=& \dot{\a}\dot{\z}-\a'\z'
     \nonumber \\
     &\rightarrow& \w<\a>\z'
 \label{aaa} \err
and
\brr \der_{[+}\Phi_1\der_{-]}\Phi_2| &=& \dot{\a}\z'-\a'\dot{\z}
     \nonumber \\
     &\rightarrow& \w<\a>\dot{\z},
 \label{bbb} \err
where $A_{(+}B_{-)}=A_+B_-+A_-B_+$, $A_{[+}B_{-]}=A_+B_--A_-B_+$,
the arrow implies that $\a\rightarrow<\a>=\exp{(-\w|\tau-\tau_0|)}$,
$\chi_\pm\rightarrow 0$ and $A\rightarrow 0$, and
$|$ indicates the lowest component of the indicated superfield
expression.  Expressions (\ref{aaa}) and (\ref{bbb}) are useful when used in
conjunction with the following facts.  If $\a\rightarrow<\a>,
\chi_\pm\rightarrow 0$, and $A\rightarrow 0$, then
\brr & & \int d\t^+d\t^-
     F(\dot{\Phi}_1,\Phi_1';\Phi_2,\dot{\Phi}_2,\Phi_2')
     D_{(+}\Phi_1D_{-)}\Phi_2 \nonumber \\
     & & \hspace{.3in}
     \rightarrow \w<\a>\Biggl\{-F|\dot{\z}+\frac{i}{\sqrt{2}}
     \frac{\d F}{\d\dot{\Phi}_1}|
     (\psi_+\dot{\psi}_++\psi_-\dot{\psi}_-) \nonumber \\
     & & \hspace{1.2in}
     +\frac{i}{\sqrt{2}}
     \frac{\d F}{\d\dot{\Phi}_1}|
     (\psi_+\dot{\psi}_++\psi_-\dot{\psi}_-)\Biggr\}
 \label{fguy} \err
and
\brr & & \int d\t^+d\t^-
     G(\dot{\Phi}_1,\Phi_1';\Phi_2,\dot{\Phi}_2,\Phi_2')
     D_+\Phi_2D_-\Phi_2 \nonumber \\
     & & \hspace{.3in}
     \rightarrow -\hf \w^2<\a>^2G|.
 \label{gguy} \err
Note that (\ref{gguy}) only involves $<\a(\tau)>, \dot{\z}$ and $\z'$
when $\Phi_2$ is replaced by its VEV.  Notice also that the fermionic
part of the right hand side of (\ref{fguy}) has the same structure
as the fermionic part of (\ref{laglin}).  We can thus use (\ref{fguy})
and (\ref{gguy})
to find the correct functions $F$ and $G$ to reproduce the fermionic
part of
the order $\gbig(\tau)$
collective field Lagrangian. Implementing this procedure, we see
immediately, from comparing (\ref{laglin}) and (\ref{fguy}), that we
require
\brr \frac{\d F}{\d \dot{\Phi}_1}| &\rightarrow& f(<\a>)\z'
     \nonumber \\
     \frac{\d F}{\d\Phi_1'}| &\rightarrow& f(<\a>)\dot{\z},
 \err
where $f(<\a>)=f(\exp{(-\w|\tau-\tau_0|)})=\gbig(\tau)$.  By comparison
with
(\ref{gdef}), we find that
\bq f(\a)=4\sqrt{\pi}g\frac{\frac{1}{\k}\a^2}
         {(1-\frac{1}{\k}\a^2)^2},
 \label{donner} \eq
where $\k$ is defined in (\ref{kdef}).
Using (\ref{aaa}) and (\ref{bbb}), an appropriate function, $F$, is
immediately seen to be
\bq F=\frac{f(\Phi_2)}{\w^3\Phi_2^3}
    \der_{(+}\Phi_1\der_{-)}\Phi_2\der_{[+}\Phi_1\der_{-]}\Phi_2.
 \label{ffoo} \eq
Thus
\brr & & \int d\t^+d\t^-FD_{(+}\Phi_1D_{-)}\Phi_2 \nonumber \\
     & & \hspace{.3in} \rightarrow
     -f(\a)\dot{\z}^2\z'
     +\frac{i}{\sqrt{2}}\z'(\psi_+\dot{\psi}_++\psi_-\dot{\psi}_-)
     \nonumber \\
     & & \hspace{1.3in}
     +\frac{i}{\sqrt{2}}\dot{\z}(\psi_+\psi_+'+\psi_-\psi_-').
 \err
By construction, the fermionic part of this expression reproduces the
fermionic part of the collective field Lagrangian at first order in
$\gbig(\tau)$.  In order that we also match the bosonic part of the
order $\gbig(\tau)$
collective field Lagrangian, we must add to this result
another supersymmetric expression to supply the difference,
which is
\bq f(<\a>)\{-\frac{1}{6}\z^{'3}+\hf\dot{\z}^2\z'\}.
 \eq
Accordingly, we use (\ref{gguy}) to determine the function $G$, which
is found to be
\bq G=\frac{f(\Phi_2)}{\w^5\Phi_2^5}
    \Biggl\{\frac{1}{3}(\der_{(+}\Phi_1\der_{-)}\Phi_2)^3
    -(\der_{[+}\Phi_1\der_{-]}\Phi_2)^2\der_{(+}\Phi_1\der_{-)}\Phi_2\Biggr\}.
 \label{ggoo} \eq
The complete supersymmetrization of the first order interactions in
the collective field Lagrangian is then
given by combining (\ref{fguy}) and (\ref{gguy}), where functions
$F$ and $G$ are specified in (\ref{ffoo}) and (\ref{ggoo}) respectively.
The result is
\brr \L_1^{(eff)} &=& \int d\t^+d\t^-\Biggl\{
     \frac{f(\Phi_2)}{\w^3\Phi_2^3}
     \der_{(+}\Phi_1\der_{-)}\Phi_2\der_{[+}\Phi_1\der_{-]}\Phi_2
     D_{(+}\Phi_1D_{-)}\Phi_2
     \nonumber \\
     & & \hspace{.3in}
     +\frac{1}{3}\frac{f(\Phi_2)}{\w^5\Phi_2^5}
     (\der_{(+}\Phi_1\der_{-)}\Phi_2)^3
     D_+\Phi_2D_-\Phi_2
     \nonumber \\
     & & \hspace{.3in}
     -\frac{f(\Phi_2)}{\w^5\Phi_2^5}
     (\der_{[+}\Phi_1\der_{-]}\Phi_2)^2\der_{(+}
     \Phi_1\der_{-)}\Phi_2
     D_+\Phi_2D_-\Phi_2\Biggr\}
 \err
where function $f$ is defined in (\ref{donner}).
Adding this result to (\ref{lago1}) and (\ref{lago2}), the supersymmetric
effective theory, up to first order in $\gbig(\tau)$, is
\brr \L^{(eff)} &=&
    \int d\t^+d\t^-\Biggl\{ D_+\Phi_1D_-\Phi_1 \nonumber \\
    & & \hspace{.3in}
    +F_1(\Phi_2)D_+\Phi_2D_-\Phi_2
    -\frac{1}{\w^2}F_2(\Phi_2)\der_-D_+\Phi_2\der_+D_-\Phi_2 \nonumber \\
    & & \hspace{.3in}
    +\frac{f(\Phi_2)}{\w^3\Phi_2^3}
     \der_{(+}\Phi_1\der_{-)}\Phi_2\der_{[+}\Phi_1\der_{-]}\Phi_2
     D_{(+}\Phi_1D_{-)}\Phi_2
     \nonumber \\
     & & \hspace{.3in}
     +\frac{1}{3}\frac{f(\Phi_2)}{\w^5\Phi_2^5}
     (\der_{(+}\Phi_1\der_{-)}\Phi_2)^3
     D_+\Phi_2D_-\Phi_2
     \nonumber \\
     & & \hspace{.3in}
     -\frac{f(\Phi_2)}{\w^5\Phi_2^5}
     (\der_{[+}\Phi_1\der_{-]}\Phi_2)^2\der_{(+}
     \Phi_1\der_{-)}\Phi_2
     D_+\Phi_2D_-\Phi_2\Biggr\}.
 \label{lag1} \err
In terms of component fields, this becomes
\brr \L^{(eff)} &=& +\der_+\z\der_-\z
     +Z^2
     -i\psi_+\der_-\psi_+
     -i\psi_-\der_+\psi_- \nonumber\\
    & & +F_1(\a)\der_+\a\der_-\a-\frac{1}{\w^2}(\der_+\der_-\a)^2
    \nonumber \\
    & & +F_1(\a)A^2-\frac{1}{\w^2}F_2(\a)\der_+A\der_-A \nonumber \\
    & & -iF_1(\a)\chi_+\der_-\chi_+
        -\frac{i}{\w^2}F_2(\a)\der_-\chi_+\der_+\der_-\chi_+
        \nonumber \\
    & & -iF_1(\a)\chi_-\der_+\chi_-
        -\frac{i}{\w^2}F_2(\a)\der_+\chi_-\der_-\der_+\chi_-
        \nonumber \\
    & & +\sum_n{\cal{O}}(\a^n\chi_+\chi_-+\a^{n-1}A\chi_+\chi_-).
     \nonumber \\
     & & -\hf{f(\a)\dot{\z}^2\z'}
     -\frac{1}{6}f(\a)\z^{'3}\nonumber \\
     & & +\frac{i}{\sqrt{2}}{f(\a)\z'}
     (\psi_+\dot{\psi}_++\psi_-\dot{\psi}_-)
     +\frac{i}{\sqrt{2}}
     {f(\a)\dot{\z}}
     (\psi_+\psi_+^{'}+\psi_-\psi_-^{'})
     \nonumber \\
     & &
     +\O\Biggl\{\der\z(\psi\chi+\chi\chi+Z\psi\chi+Z\chi\chi
          +A\psi\psi+A\psi\chi)+\psi\psi\chi+\psi\chi\chi\Biggr\}
 \label{compost} \err
In the above analysis we have implicitly assumed that
$<\a>=\exp{(-\w|\tau-\tau_0|)},<\chi_\pm>=<A>=0$
and $<Z>=0$ remain solutions
of the equations of motion at the order $\gbig(\tau)$ level.  We now show
that this assumption is indeed correct.
The $\a$ equation of motion is given by
(\ref{mot1}) modified by terms of order $A^2$, order $\chi^2$, and
of order $\der\z$.  By construction, the functions $F_1$ and $F_2$
ensure that (\ref{mot1}) is satisfied by the exponential solution.
If $<A>=<\chi_\pm>=0$ and if $<\z>=$constant, then the $\a$
equation is still satisfied by the same exponential.
The fields $\chi_\pm, A$ and
$Z$ all occur at least bilinearly or coupled to $\der\z$ in
the component Lagrangian (\ref{compost}). Also, the field $\z$
always appears with derivatives
and at least quadratically.  It is clear then
that the following is a solution to the set of component field
equations derived from (\ref{compost}),
\brr <\a> &=& \exp{(-\w|\tau-\tau_0|)} \nonumber \\
     <\z> &=& constant \nonumber \\
     <\chi_\pm> &=& 0 \nonumber \\
     <\psi_\pm> &=& 0 \nonumber \\
     <A> &=& 0 \nonumber \\
     <Z> &=& 0.
 \label{solon} \err
To expose the relation of the component Lagrangian (\ref{compost})
to the collective field Lagrangian, we replace the fields
$\a, \chi_\pm$, and $A$ with the VEV's listed above.
We replace the fields $\z$ and $\psi_\pm$ with fields shifted
around their VEV's.  But since $<\psi_\pm>=0$ and $<\der\z>=0$,
the shifted fields appear coupled precisely as do the unshifted
fields.  For this reason we do not distinguish, notationally,
the fields $\z$ and $\psi_\pm$ in the effective Lagrangian
(\ref{compost}) from the ``shifted" fields appearing in the collective
field Lagrangian (\ref{lagcollfinal}).  The field $Z$ is auxiliary.
Since it always appears either bilinearly or coupled to $\chi_\pm$
it follows, since we do not exhibit the fluctuation of $\chi_\pm$
around its vanishing VEV, that the auxiliary field $Z$ is of no
consequence to the shifted Lagrangian.  Implementing the
process of replacing fields with VEV's and shifting $\z$ and $\psi_\pm$
around their VEV's, we recover, from the supersymmetric Lagrangian
(\ref{compost}), the collective field Lagrangian (\ref{lagcollfinal})
to first order in $\gbig(\tau)$.
{\fl{\it c) All orders:}}

The procedure outlined above may be carried out to all orders in the
coupling $\gbig(\tau)$.   It is straightforward to do this, so we will simply
quote the result. This is, however,  a non-trivial statement. The Lagrangian is
highly non-linear and we find it remarkable that it in fact exists.
The all orders supersymmetric effective theory is
\brr \L^{(eff)} &=&
     \int d\t_+d\t_-\Biggl\{
     D_+\Phi_1D_-\Phi_1  \nonumber \\
     & & \hspace{.6in}
     +F_1(\Phi_2)D_+\Phi_2D_-\Phi_2
    -\frac{1}{\w^2}F_2(\Phi_2)\der_-D_+\Phi_2\der_+D_-\Phi_2 \nonumber \\
     & & \hspace{.6in}
     +\frac{f(\Phi_2)}{\w^3\Phi_2^3}
     \frac{\der_{(+}\Phi_1\der_{-)}\Phi_2
           \der_{[+}\Phi_1\der_{-]}\Phi_2}
          {1+\frac{f(\Phi_2)}{\w\Phi_2}
           \der_{(+}\Phi_1\der_{-)}\Phi_2}
           D_{(+}\Phi_1D_{-)}\Phi_2 \nonumber \\
     & & \hspace{.6in}
     +\frac{1}{3}\frac{f(\Phi_2)}{\w^5\Phi_2^5}
     (\der_{[+}\Phi_1\der_{-]}\Phi_2)^3
     D_+\Phi_2D_-\Phi_2 \nonumber \\
     & & \hspace{.6in}
     -\frac{f(\Phi_2)}{\w^5\Phi_2^5}
     \frac{(\der_{[+}\Phi_1\der_{-]}\Phi_2)^2
     \der_{(+}\Phi_1\der_{-)}\Phi_2}
     {1+\frac{f(\Phi_2)}{\w\Phi_2}
     \der_{(+}\Phi_1\der_{-)}\Phi_2}
     D_+\Phi_2D_-\Phi_2 \Biggr\}.
 \label{enddd} \err
In terms of component fields, this reads
\brr \L^{(eff)} &=& +\der_+\z\der_-\z
     +Z^2
     -i\psi_+\der_-\psi_+
     -i\psi_-\der_+\psi_- \nonumber\\
    & & +F_1(\a)\der_+\a\der_-\a
         -\frac{1}{\w^2}F_2(\a)(\der_+\der_-\a)^2
    \nonumber \\
    & & +F_1(\a)A^2-\frac{1}{\w^2}F_2(\a)\der_+A\der_-A \nonumber \\
    & & -iF_1(\a)\chi_+\der_-\chi_+
        -\frac{i}{\w^2}F_2(\a)\der_-\chi_+\der_+\der_-\chi_+
        \nonumber \\
    & & -iF_1(\a)\chi_-\der_+\chi_-
        -\frac{i}{\w^2}F_2(\a)\der_+\chi_-\der_-\der_+\chi_-
        \nonumber \\
    & & +\sum_n{\cal{O}}(\a^n\chi_+\chi_-+\a^{n-1}A\chi_+\chi_-).
     \nonumber \\
     & & -\hf\frac{f(\a)\dot{\z}^2\z'}{1+f(\a)\z'}
     -\frac{1}{6}f(\a)\z^{'3}\nonumber \\
     & & +\frac{i}{\sqrt{2}}\frac{f(\a)\z'}{1+f(\a)\z'}
     (\psi_+\dot{\psi}_++\psi_-\dot{\psi}_-)
     +\frac{i}{\sqrt{2}}
     \frac{f(\a)\dot{\z}}{[1+f(\a)\z']^2}
     (\psi_+\psi_+^{'}+\psi_-\psi_-^{'})
     \nonumber \\
     & &
     +\O\Biggl\{\der\z(\psi\chi+\chi\chi+Z\psi\chi+Z\chi\chi
          +A\psi\psi+A\psi\chi)+\psi\psi\chi+\psi\chi\chi\Biggr\}
  \err
The component field equations of motion derived from (\ref{enddd})
are of the same type as those derived from the first order
Lagrangian (\ref{compost}) and the solution
(\ref{solon}) satisfies these equations as well.  By replacing
all fields by their VEV's and then shifting $\z$ and $\psi_\pm$
around their VEV's, we recover, from the supersymmetric Lagrangian
(\ref{enddd}), the $\tau$-dependent collective field Lagrangian
(\ref{lagcollfinal}). This, then, is the supersymmetric effective theory
which we had set out to construct.
This is the essential result of this paper.

Since the Lagrangian (\ref{enddd}) is Poincare invariant and
supersymmetric, there exists a more general class of solutions
to the field equations than those given in (\ref{solon}).
The more general solution is given by
\brr <\a> &=& \exp\Biggl\{
     \w[|t-t_0|\sinh\t_0-|\tau-\tau_0|\cosh\t_0]\Biggr\}
     \nonumber \\
     <\z> &=& constant \nonumber \\
     <\chi_\pm> &=& \n_0^\pm<\a>  \nonumber \\
     <\psi_\pm> &=& 0 \nonumber \\
     <A> &=& 0 \nonumber \\
     <Z> &=& 0,
 \label{solongen} \err
where $\t_0$ is a Lorentz zero mode, $t_0$ and $\tau_0$ are
translational zero modes, and $\n_0^\pm$ are supersymmetric
zero modes.  The solution (\ref{solon}) corresponds to the
choice $\t_0=\n_0^\pm=0$.

The results derived in this section are sufficiently complicated
that, for clarity, we will now recapitulate them.  The main
result is the (1,1) supersymmetric effective Lagrangian,
\brr \L^{(eff)} &=&
     \int d\t_+d\t_-\Biggl\{
     D_+\Phi_1D_-\Phi_1  \nonumber \\
     & & \hspace{.6in}
     +F_1(\Phi_2)D_+\Phi_2D_-\Phi_2
    -\frac{1}{\w^2}F_2(\Phi_2)\der_-D_+\Phi_2\der_+D_-\Phi_2 \nonumber \\
     & & \hspace{.6in}
     -\frac{f(\Phi_2)}{\w^3\Phi_2^3}
     \frac{\der_{(+}\Phi_1\der_{-)}\Phi_2
           \der_{[+}\Phi_1\der_{-]}\Phi_2}
          {1+\frac{f(\Phi_2)}{\w\Phi_2}
           \der_{(+}\Phi_1\der_{-)}\Phi_2}
           D_{(+}\Phi_1D_{-)}\Phi_2 \nonumber \\
     & & \hspace{.6in}
     +\frac{1}{3}\frac{f(\Phi_2)}{\w^5\Phi_2^5}
     (\der_{[+}\Phi_1\der_{-]}\Phi_2)^3
     D_+\Phi_2D_-\Phi_2 \nonumber \\
     & & \hspace{.6in}
     -\frac{f(\Phi_2)}{\w^5\Phi_2^5}
     \frac{(\der_{[+}\Phi_1\der_{-]}\Phi_2)^2
     \der_{(+}\Phi_1\der_{-)}\Phi_2}
     {1+\frac{f(\Phi_2)}{\w\Phi_2}
     \der_{(+}\Phi_1\der_{-)}\Phi_2}
     D_+\Phi_2D_-\Phi_2 \Biggr\},
 \label{efflag} \err
where
\brr F_1(\Phi_2) &=& -\frac{1}{48\pi\k\w^2 g^2}
     (\frac{11}{5}\frac{\k^3}{\Phi_2^6}
     -\frac{28}{3}\frac{\k^2}{\Phi_2^4}
     +18\frac{\k}{\Phi_2^2}
     -4
     +\frac{5}{3}\frac{\Phi_2^2}{\k}) \nonumber \\
     F_2(\Phi_2) &=&  -\frac{1}{48\pi\k\w^2 g^2}
     (-\frac{2}{5}\frac{\k^3}{\Phi_2^6}
     +\frac{8}{3}\frac{\k^2}{\Phi_2^4}
     -12\frac{\k}{\Phi_2^2}
     -8
     +\frac{2}{3}\frac{\Phi_2^2}{\k}) \nonumber \\
     f(\Phi_2) &=& 4\sqrt{\pi}g\frac{\frac{1}{\k}\Phi_2^2}
         {(1-\frac{1}{\k}\Phi_2^2)^2}.
 \err
Note that the effective Lagrangian has three independent parameters,
$\w, g$ and $\k$.  The parameter $\w$ is a mass,
$g$ is an inverse mass, and $\k$ is a dimensionless number.
The equations of motion derived from (\ref{efflag}) are satisfied
by the solution (\ref{solongen}).  This solution is labeled by the
translational zero modes $t_0$ and $\tau_0$, a Lorentz zero mode
$\t_0$ and by supersymmetric zero modes $\n_0^\pm$.  In a preferred
frame of reference, $\t_0=\n_0^+=\n_0^-=0$, and the solution
(\ref{solongen}) becomes equivalent to (\ref{solon}).  If we
substitute the solution (\ref{solon}) into (\ref{efflag}),
thus freezing the ``heavy" fields $\a, \chi_+$ and $\chi_-$ at their
VEV's, and shift the light fields $\z, \psi_-$ and $\psi_+$
around their VEV's, we recover the collective field Lagrangian
(\ref{lagcollfinal})
derived from the $d=1, \cal{N}=2$ supersymmetric matrix
model.
In this expression the coupling parameter $\gbig(\tau)=f(<\a>)$
is given by (\ref{ggg}) and (\ref{gdef}), and is plotted in figure 4.
Notice that this coupling parameter blows up at the boundaries
of a region centered at $\tau=\tau_0$ with width
$\s=\frac{1}{\w}\ln\k$.
Outside of this region, the high density collective field theory
(\ref{lagcollfinal}) is valid.
Within the
region $|\tau-\tau_0|<\s/2$ or, equivalently, in the region
$|x|<1/(\w\sqrt{g})$
however, the
collective field theory must describe a finite number of eigenvalues.
In this case, the appropriate form for the collective field theory
is given in (\ref{lagcollpp}).  This Lagrangian is completely
equivalent to the original eigenvalue Lagrangian (\ref{lageigsusy})
or (\ref{lageig}), which is, in most cases, easier to use.  We now turn
to a discussion of instanton-like solutions to the Euclidean
equations of motion of this low density eigenvalue theory.

\renewcommand{\theequation}{6.\arabic{equation}}
\setcounter{equation}{0}
{\fl{\bf 6. Eigenvalue Instantons}}

In this section we will construct solutions to the Euclidean field
equations derived from the low density eigenvalue Lagrangian
(\ref{lageigsusy}).
The Euclideanized version
of Lagrangian (\ref{lageigsusy}), is given by
\bq L_E=\sum_i\{\hf\dot{\l}_i^2+\hf(\frac{\der W_{eff}}{\der\l_i})^2
    +\frac{i}{2}\chi_{1i}\dot{\chi}_{1i}
    -\frac{i}{2}\chi_{2i}\dot{\chi}_{2i}\}
    +i\sum_{ij}\chi_{1i}\chi_{2j}
    \frac{\der^2W_{eff}}{\der\l_i\der\l_j},
   \eq
where the dot means differentiation with respect to Euclidean time
$\t$.
In this expression,
\bq W_{eff}(\l)=W(\l)+w(\l),
 \eq
where $W(\l)$ is the superpotential, and $w(\l)$ is the modification,
given in (\ref{dasher}), which results from the restriction of the
underlying matrix model to its singlet sector.  In the static
ground states discussed above no eigenvalues populate the low
density region.  In this section we will describe additional
solutions to the Euclidean field equations in which only one
eigenvalue populates the low density region.  The modification to
the superpotential, $w(\l)$, induces only a local inter-eigenvalue force.
Therefore,
if only a single eigenvalue exists in the low density region, we can
neglect $w(\l)$ in the Lagrangian.
The dynamics of this single eigenvalue
and its fermionic superpartners is then described by the following
Euclideanized Lagrangian,
\bq L_E=\hf\dot{\l}^2
    +\hf(\frac{\der W}{\der\l})^2
    -\frac{i}{2}\chi_1\dot{\chi}_1
    +\frac{i}{2}\chi_2\dot{\chi}_2
    +i\chi_1\chi_2\frac{\der^2W}{\der\l^2}.
 \label{lageuc}\eq
This Lagrangian in symmetric under the Euclidean
$d=1, {\cal{N}}=2$
supersymmetry transformation,
\brr \d\l &=& i\n^1\chi_1+i\n^2\chi_2 \nonumber \\
     \d\chi_1 &=& +\n^1\dot{\l}-\n^2W' \nonumber \\
     \d\chi_2 &=& -\n_2\dot{\l}+\n^1W'
 \err
The Euclidean field equations derived from (\ref{lageuc}) are
\brr \ddot{\l}-W'W'' &=& 0 \nonumber \\
     \dot{\chi}_1-W''\chi_2 &=& 0 \nonumber \\
     \dot{\chi}_2-W''\chi_1 &=& 0,
 \label{euck} \err
where $W'=\der W/\der\l$ and
$W''=\der^2W/\der\l^2$.  Now, recall that the superpotential
depends on $N$, the total number of eigenvalues in the matrix
model.  Specifically, from (\ref{potform}), we have
\bq W(\l)=Nc_0+\sqrt{N}c_1\l+\frac{1}{6}\frac{c_3}{\sqrt{N}}\l^3,
 \eq
where $c_0, c_1$ and $c_3$ are arbitrary finite constants.
It follows that (\ref{euck}) become
\brr \ddot{\l}+\w^2 &=& \hf a^2\l^3 \nonumber \\
     \dot{\chi}_1 &=& a\l\chi_2 \nonumber \\
     \dot{\chi}_2 &=& a\l\chi_1,
 \label{pudding}\err
where $\w^2=-c_1c_3>0$ and $a=c_3/\sqrt{N}$.
Since $N$ is very large, $a$ is very small.  Thus, for finite
$N$, the field equations (\ref{pudding}) are only slight perturbations
from the following system,
\brr \dot{\l}+\w^2\l &=& 0 \nonumber \\
     \dot{\chi}_1 &=& 0 \nonumber \\
     \dot{\chi}_2 &=& 0.
 \label{jello}\err
In the limit $N\rightarrow\infty$, these equations become exact.
The general solution to the single-eigenvalue Euclidean
field equations in the large $N$ limit is then
\brr \l^* &=& A\sin\{\w(\t-\t_0)\}
             +B\cos\{\w(\t-\t_0)\} \nonumber \\
     \chi_1^* &=& -\n_{10} \nonumber \\
     \chi_2^* &=& -\n_{20},
 \label{blek} \err
where $A, B$ and
$\t_0$ are real commuting constants and $\n_{10}$ and
$\n_{20}$ are real anticommuting constants.
We want to consider, for reasons to be discussed below,
solutions, which we denote by $\l^{(+)}$ and $\chi_{1,2}^{(+)}$,
satisfying
the following boundary condition,
\brr \l^{(+)}|_{\t=\t_0\mp\pi/(2\w)} &=&
     \frac{\mp 1}{\w\sqrt{g}} \nonumber \\
     \dot{\l}^{(+)}|_{\t=\t_0\mp\pi/(2\w)} &=& 0.
 \err
It follows from (\ref{blek}) that
\brr  \l^{(+)} &=& \frac{-1}{\w\sqrt{g}}\sin\{\w(\t-\t_0^{(+)})\}
      \nonumber \\
      \chi_1^{(+)} &=& -\n_{10}^{(+)} \nonumber \\
      \chi_2^{(+)} &=& -\n_{20}^{(+)}.
 \label{instplus} \err
Similarly, we consider solutions $\l^{(-)}$ and $\chi_{1,2}^{(-)}$
which satisfy the boundary condition,
\brr \l^{(-)}|_{\t=\t_0\mp\pi/(2\w)} &=&
     \frac{\pm 1}{\w\sqrt{g}} \nonumber \\
     \dot{\l}^{(-)}|_{\t=\t_0\mp\pi/(2\w)} &=& 0.
 \err
Thus,
\brr  \l^{(-)} &=& \frac{+1}{\w\sqrt{g}}\sin\{\w(\t-\t_0^{(-)})\}
      \nonumber \\
      \chi_1^{(-)} &=& -\n_{10}^{(-)} \nonumber \\
      \chi_2^{(-)} &=& -\n_{20}^{(-)}.
 \label{instminus} \err
The parameters $\t_0^{(\pm)}, \n_{10}^{(\pm)}$ and
$\n_{20}^{(\pm)}$ are zero-modes associated with these solutions.
It is very enlightening to reexpress these results
in the language of collective field theory.
To do this, we use the definitions (\ref{colldef}),
which for the single eigenvalue case in Euclidean time are
\brr \vphi(x,\t) &=& \Theta(x-\l(\t)) \nonumber \\
     \psi_1(x,\t) &=& \d(x-\l(\t))\chi_{1}(\t) \nonumber \\
     \psi_2(x,\t) &=& \d(x-\l(\t))\chi_{2}(\t).
 \label{eigcoll} \err
Recall that in the high density region we let
$\vphi=\tilde{\vphi}_0(x)+\frac{1}{\sqrt{\pi}}\z$,
where $\tilde{\vphi}_0(x)$ was the vacuum solution given in
(\ref{phix}).  Here, in the low density region, we will also
express $\vphi$ as $\vphi=\tilde{\vphi}_0(x)+\frac{1}{\sqrt{\pi}}\z$.
Now, however, $\tilde{\vphi}_0(x)=0$ and, hence,
$\vphi=\frac{1}{\sqrt{\pi}}\z$.  Substituting solutions
(\ref{instplus}) and (\ref{instminus}) into (\ref{eigcoll})
yields the collective field theory vacuum configurations
\brr \z^{(\pm)}(x,\t) &=&
     \sqrt{\pi}\Theta\bigg{(} x\mp\frac{1}{\w\sqrt{g}}
     \sin[\w(\t-\t_0^{(\pm)})]\bigg{)}
     \nonumber \\
     \psi_1^{(\pm)}(x,\t) &=&
     \d\bigg{(}  x \mp\frac{1}{\w\sqrt{g}}
     \sin[\w(\t-\t_0^{(\pm)})]\bigg{)}\n_{10}^{(\pm)}
     \nonumber \\
     \psi_2^{(\pm)}(x,\t) &=&
     \d\bigg{(}x\mp\frac{1}{\w\sqrt{g}}
     \sin[\w(\t-\t_0^{(\pm)})]\bigg{)}\n_{20}^{(\pm)}.
 \label{xinstantons}) \err
These expressions are valid over the region
$-1/(\w\sqrt{g})<x<+1/(\w\sqrt{g})$.  Outside of this region,
for $|x|\ge 1/(\w\sqrt{g})$ (or, equivalently, for
$|\tau|\ge \s/2$), the high density collective field theory
is valid and the vacuum solution is given by (\ref{solon}).
Thus, for $|x|\ge 1/(\w\sqrt{g})$, we take $\z^{(\pm)}=$constant
and $\psi_1^{(\pm)}=\psi_2^{(\pm)}=0$.  We can then match these
vacuum solutions at both $x=-1/(\w\sqrt{g})$ and at
$x=+1/(\w\sqrt{g})$, and therefore extend the configurations
(\ref{xinstantons}) over all of space.  This requires that,
for $x<-1/(\w\sqrt{g})$, we choose
$\z^{(\pm)}=\psi_1^{(\pm)}=\psi_2^{(\pm)}=0$, and, for
$x>+1/(\w\sqrt{g})$, we choose $\z^{(\pm)}=\sqrt{\pi}$
and $\psi_1^{(\pm)}=\psi_2^{(\pm)}=0$.  The $(+)$ configurations are
depicted in figure 5.
The configurations
$\z^{(\pm)}$ are kinks which move across the low density
region in Euclidean time.  Since they are kinks, these configurations
are topologically stable.
The $\z^{(\pm)}$ describe an eigenvalue
moving in Euclidean time from $x=\mp 1/\sqrt{\w^2g}$
at $\t=\t_0-\pi/(2\w)$ to
$x=\pm 1/\sqrt{\w^2g}$ at $\t=\t_0+\pi/(2\w)$.
This represents a quantum mechanical tunneling
of an eigenvalue across the low density region.  For the cases
where $\n_{10}^{(\pm)}\ne 0$ or $\n_{20}^{(\pm)}\ne 0$ the
eigenvalues are accompanied by fermions which also tunnel
across the low density region at the same time.
Notice that configurations
with a superscript $(+)$
represent tunneling from left to right and that configurations
with a superscript $(-)$ represent tunneling from right to left.

\ \vspace{2pc}

\centerline{\tenrm Figure~5. Right-moving instantons}

Now recall from (\ref{xoft}) that in the low density region
$|x|<1/(\w\sqrt{g})$, we have $x=\frac{1}{\w\sqrt{g}}
\sin\{\frac{\pi}{\s}(\tau-\tau_0)\}$. Using this transformation,
we can represent the above vacuum in $\tau$ space.
Thus,
\brr \z^{(\pm)}(\tau,\t) &=&
     \sqrt{\pi}\Theta\bigg{(}\frac{1}{\w\sqrt{g}}
     [\sin\frac{\pi}{\s}(\tau-\tau_0)\mp\sin(\t-\t_0^{(\pm)})]\bigg{)}
     \nonumber \\
     \psi_1^{(\pm)}(\tau,\t) &=&
     \d\bigg{(}\frac{1}{\w\sqrt{g}}
     [\sin\frac{\pi}{\s}(\tau-\tau_0)\mp\sin(\t-\t_0^{(\pm)})]\bigg{)}
     \n_{10}^{(\pm)} \nonumber \\
     \psi_2^{(\pm)}(\tau,\t) &=&
     \d\bigg{(}\frac{1}{\w\sqrt{g}}
     [\sin\frac{\pi}{\s}(\tau-\tau_0)\mp\sin(\t-\t_0^{(\pm)})]\bigg{)}
     \n_{20}^{(\pm)}.
 \err
These expressions are valid for $(\tau_0-\s/2)<\tau<(\tau_0+\s/2)$.
As discussed above, outside this region, for $\tau<(\tau_0-\s/2)$, we
take $\z^{(\pm)}=\psi_1^{(\pm)}=\psi_2^{(\pm)}=0$, and for
$\tau>(\tau_0+\s/2)$, we take $\z^{(\pm)}=\sqrt{\pi}$ and
$\psi_1^{(\pm)}=\psi_2^{(\pm)}=0$.
The instanton
background is thus defined over all of $(\t,\tau)$ space.
One eigenvalue instantons were also discussed in a different context
in \cite{leemend}.

It is tempting to conjecture that the instantons described above
actually break the two-dimensional (1,1) supersymmetry of the
effective action.  The reasons for this speculation are the
following.  The quantum mechanical instantons are expected to break
the $d=1, \cal{N}=2$ supersymmetry.  This is a known phenomenon
in supersymmetric quantum mechanics\cite{witten,salomonson}.
The $d=2$, $(1,1)$ supersymmetry is apparently a consequence of the
$d=1, \cal{N}=2$ supersymmetry.
The exact connection
between the two supersymmetries is not well understood, however.
It is thus reasonable to assume that, if the
underlying $d=1, \cal{N}=2$ is
broken, so will be the $d=2$, $(1,1)$ supersymmetry of the effective theory.

The question of whether supersymmetry breaking indeed occurs in the
effective theory and whether it is due to single eigenvalue instantons is
currently under active investigation\cite{us}.

{\fl{\bf 7. Conclusions}}

We have derived a two-dimensional (1,1)
supersymmetric effective Lagrangian which reduces to the collective
field Lagrangian describing the most general $d=1, \cal{N}=2$
supersymmetric matrix model in the large $N$ limit,
when certain ``heavy" fields are frozen at their VEV's.
Additionally, we have shown that the dynamics of the light fields in the
effective theory include a space-dependent coupling parameter
which blows up at finite points and delineates a special zone
in which quantum mechanical, not field theoretical, considerations
need to be incorporated into the physical picture of the system.
The quantum mechanical aspect of the effective field theory
relates to individual eigenvalue dynamics of the underlying matrix model.
It is a ``stringy" aspect of the effective theory.
We have also indicated how these eigenvalue instantons might induce
supersymmetry breaking in the effective field theory.

\renewcommand{\theequation}{A.\arabic{equation}}
\setcounter{equation}{0}
{\fl{\bf Appendix A: Derivation of the Effective Singlet Sector
Lagrangian for
the Quantum Bosonic Matrix Model }}

In this Appendix we provide a discussion of some results cited in
section 2.  Specifically, what follows is a detailed calculation
of the effective $U(N)$ singlet sector
Lagrangian relevant to the quantum
mechanical bosonic matrix model.

The Hamiltonian for a bosonic matrix model is
\bq H=\hf Tr\Pi^2_M + V(M),
 \label{h} \eq
where
\bq V(M)=\sum_n a_n Tr M^n
 \eq
and $M$ is a time dependent Hermitian $N\times N$ matrix.
Canonically quantizing, we replace $M_{ij}$ with the operator
$\hat{M}_{ij}$ and $\Pi_{M_{ij}}$ with
$\hat{\Pi}_{M_{ij}}$.  We work in the $M$ basis, where
$\hat{\Pi}_{M_{ij}}=-i{\der}/{\der{M}_{ij}}$.  We would like to
express the quantum operator Hamiltonian, $\hat{H}$, in terms of
the matrix eigenvalues, ${\l}_i$ and their conjugate
momenta, $\hat{\Pi}_{\l_i}=-i\der/\der{\l}_i$.
Toward this goal we embark on the following discussion.
Given a unitary matrix, $U_{ij}$,
\bq \sum_kU^\dagger_{ik}U_{kj}=\d_{ij},
 \label{udel} \eq
 we have
the relation
\bq \frac{\der U^\dagger_{ij}}{\der U_{kl}}=-U^\dagger_{ik}
    U^\dagger_{lj}.
 \label{ugh} \eq
If $M_{ij}$ is a Hermitian matrix, then there exists a $U_{ij}$
such that the unitary transformation,
\bq M_{ij}=\sum_k U^\dagger_{ik}\l_k U_{kj},
 \label{oog} \eq
relates $M_{ij}$ to the the diagonal matrix
$\l_{ij}=\l_i\d_{ij}$ which consists of the eigenvalues
of $M_{ij}$.  Given (\ref{ugh}) and (\ref{oog}), it it readily found
that
\bq \{\frac{\der M_{ij}}{\der\l_l}\}^{-1}=U^\dagger_{jl}U_{li}
 \eq
and
\bq \{\frac{\der M_{ij}}{\der U_{ls}}\}^{-1}
    =\sum_{k\ne l}\frac{U^\dagger_{jk}U_{ks}U_{li}}{\l_l-\l_k}.
 \eq
Now, using the chain rule,
\brr \frac{\der}{\der M_{ij}} &=&
         \sum_l\{\frac{\der M_{ij}}{\der\l_l}\}^{-1}\frac{\der}{\der\l_l}
         +\sum_{ls}\{\frac{\der M_{ij}}{\der U_{ls}}\}^{-1}
         \frac{\der}{\der U_{ls}} \\
    &=& \sum_l U^\dagger_{jl}U_{li}\frac{\der}{\der\l_l}
         +\sum_{ls}
         \sum_{k\ne l}\frac{U^\dagger_{jk}U_{ks}U_{li}}{\l_l-\l_k}
         \frac{\der}{\der U_{ls}}.
 \label{deem} \err
With (\ref{deem}), it is straightforward, if tedious, using (\ref{udel})
and (\ref{ugh}), to show that
\brr  \sum_i\sum_j\frac{\der}{\der M_{ij}}\frac{\der}{\der M_{ji}}
      &=& \sum_i\Biggl\{\frac{\der^2}{\der\l_i^2}
          +2\sum_{j\ne i}\frac{1}{\l_j-\l_i}\frac{\der}{\der\l_i}\Biggr\}
          \nonumber \\
      &-& \sum_i\sum_{j\ne i}\frac{1}{(\l_i-\l_j)^2}
          \Biggl\{\sum_s U_{is}\frac{\der}{\der U_{is}}
          +\sum_s\sum_t U_{is}U_{jt}\frac{\der}{\der U_{js}}
          \frac{\der}{\der U_{it}}\Biggr\}.
 \label{teed} \err
This relation also holds when $M, \l$ and $U$ are replaced with quantum
operators $\hat{M}, \hat{\l}$ and $\hat{U}$.
Our interest is in the quantum theory.
Henceforth, we restrict our attention to quantum states which have no
$U_{ij}$ dependance.  This subspace of the Hilbert space
consists of $U(N)$ singlets.
The last two terms in (\ref{teed}) annihilate this subspace and can
therefore be ignored.  We have the relations,
$\hat{\Pi}_{M_{ij}}=-i\frac{\der}{\der{M}_{ij}}$ and
$\hat{\Pi}_{\l_i}=-i\frac{\der}{\der{\l}_i}$,
so that (\ref{teed}) can now be rewritten
\bq \sum_{ij}\hf\hat{\Pi}_{M_{ij}}\hat{\Pi}_{M_{ji}}=
    \sum_i\Biggl\{\hf\hat{\Pi}^2_{\l_i}
    -i\sum_{j\ne i}\frac{1}{{\l}_i-{\l}_j}\hat{\Pi}_{\l_i}\Biggr\}
 \label{thing} \eq
Also, somewhat trivially,
\brr V(\hat{M}) &=& \sum_n a_n Tr {M}^n \\
     &=& \sum_n a_n\sum_i{\l}_i^n \\
     &=& \sum_i V({\l}_i).
 \label{v} \err
Thus, given (\ref{h}), (\ref{thing}), and (\ref{v}), the
quantum operator Hamiltonian
relevant to the $U(N)$ singlet sector of the matrix model is
\bq \hat{H}_S =\sum_i\Biggl\{\hf\hat{\Pi}^2_{\l_i}
      -i\sum_{j\ne i}\frac{1}{{\l}_i-{\l}_j}
      \hat{\Pi}_{\l_i}+V({\l}_i)\Biggr\}
 \label{trombone}\eq
The partition function is given by
\brr Z_N(a_n) &=&\int[d\Pi_{\l}][d\l]\exp
                 {i\int dt\sum_i\Biggl\{\Pi_{\l_i}\dot{\l_i}
                  -H\Biggr\}} \\
              &=&\int[d\Pi_{\l}][d\l]\exp
                 {i\int dt\sum_i\Biggl\{-\hf\Pi^2_{\l_i}
                  +(\dot{\l_i}+i\sum_{j\ne i}\frac{1}{\l_j-\l_i})
                  \Pi_{\l_i}-V(\l_i)\}}.
 \err
The $[d\Pi_{\l}]$ integration is gaussian and is easily performed.  The
result, ignoring an irrelevant constant prefactor, is
\brr Z_N(a_n) &=& \int[d\l]\exp{i\int dt\sum_i\Biggl\{
                  \hf(\dot{\l}_i+i\sum_{j\ne i}\frac{1}{\l_j-\l_i})^2
                  -V(\l_i)}\Biggr\}
                  \label{a}  \\
              &=& \int[d\l]\exp{i\int dt\sum_i\Biggl\{\hf\dot{\l}_i^2
                  -\hf(\sum_{j\ne i}\frac{1}{\l_j-\l_i})^2-V(\l_i)\Biggr\}}
                  \label{b}  \\
              &\equiv& \int[d\l]\exp{i\int dt L_S(\dot{\l},\l}).
 \err \label{c}
In passing from (\ref{a}) to (\ref{b}), we drop the cross term
because
\brr \int dt\sum_i\sum_{j\ne i}\frac{1}{\l_j-\l_i}\dot{\l}_i &=&
         -\int dt\sum_{j\ne i}\sum_{i\ne j}\frac{\dot{\l}_i-\dot{\l}_j}
          {\l_i-\l_j} \\
     &=& -\sum_{j\ne i}\sum_{i\ne j}\int dt\der_t\ln(\l_i-\l_j) \\
     &=& 0.
 \err
Thus, the $U(N)$ singlet sector of a quantum mechanical
bosonic matrix model is governed by
an effective lagrangian
\bq L_S(\dot{\l},\l)=\sum_i\Biggl\{\hf\dot{\l}_i^2-V(\l_i)
      -\hf(\sum_{j\ne i}\frac{1}{\l_j-\l_i})^2\Biggr\}.
 \label{ll} \eq
This is the result cited in section 2.

\renewcommand{\theequation}{B.\arabic{equation}}
\setcounter{equation}{0}
{\fl{\bf Appendix B: Derivation of the Supersymmetric Quantum Mechanics
as a Subsector of the Supersymmetric Quantum Matrix Model}}

In this Appendix we provide a discussion of results cited in section 3.
Specifically, what follows is a detailed extraction of the supersymmetric
quantum mechanics as a subspace of the full supersymmetric quantum
matrix model.

The Hamiltonian for an ${\cal{N}}=2$ supersymmetric matrix model is
\bq H=\sum_{ij}\Biggl\{\hf\Pi_{M_{ij}}\Pi_{M_{ji}}
     +\hf\frac{\der W(M)}{\der M_{ij}}\frac{\der W(M)}{\der M_{ij}}\Biggr\}
     +\frac{i}{2}\sum_{ijkl}[\bar{\Psi}_{ij},\Psi_{kl}]
     \frac{\der^2W(M)}{\der M_{ij}\der M_{kl}},
 \label{hambone} \eq
where
\bq W(M)=\sum_n b_nTr M^n,
 \eq
$M$ is a time-dependent commuting $N\times N$ Hermitian matrix,
$\Psi$ is an anticommuting $N\times N$ Hermitian matrix, and
$\bar{\Psi}$ is the Hermitian conjugate of $\Psi$.  Canonically
quantizing, we replace $M_{ij}$ with the operator $\hat{M}_{ij}$,
$\Pi_{M_{ij}}$ with $\hat{\Pi}_{M_{ij}}$, $\Psi_{ij}$ with
$\hat{\Psi}_{ij}$ and $\bar{\Psi}_{ij}$ with $\hat{\bar{\Psi}}$.
We also impose the following relations
\brr [\hat{\Pi}_{M_{ij}},\hat{M}_{kl}] &=& -i\d_{ik}\d_{jl} \nonumber \\
     \{\hat{\bar{\Psi}}_{ij},\hat{\Psi}_{kl}\}
     &=& \d_{ik}\d_{jl}.
 \label{rel2} \err
We henceforth
work in the $M$ basis, where $\hat{\Pi}_{M_{ij}}=-i\der/\der M_{ij}$.
The operators $\Psi$ and $\bar{\Psi}$ are annihilation and creation
operators for fermions.
We parameterize $M_{ij}$ as follows,
\bq M_{ij}=\sum_k U^\dagger_{ik}\l_k U_{kj},
  \eq
where $\l_k$ are the eigenvalues of $M_{ij}$ and $U_{ij}$ is a unitary
matrix.  This is always possible since $M$ is Hermitian.
We use the same matrix $U$ to define a ``rotated" fermion matrix,
$\chi_{ij}$,
\bq \Psi_{ij}=\sum_{kl} U^\dagger\chi_{kl}U_{lj}.
 \label{chidef} \eq
Using the chain rule, as discussed in Appendix A, it follows that
\brr \frac{\der}{\der M_{ij}} &=&
        \sum_l U^\dagger_{jl}U_{li}\frac{\der}{\der\l_l}
         +\sum_{ls}
         \sum_{k\ne l}\frac{U^\dagger_{jk}U_{ks}U_{li}}{\l_l-\l_k}
         \frac{\der}{\der U_{ls}}.
  \err
It is then straightforward to demonstrate that
\brr \frac{\der}{\der M_{ij}}\frac{\der}{\der M_{lk}}
     &=& \sum_{ab}(U_{ai}U^\dagger_{ja})(U_{bl}U^\dagger_{kb})
     \frac{\der}{\der\l_a}\frac{\der}{\der\l_b} \nonumber \\
     & & +2\sum_a\sum_{b\ne a}(U_{ai}U^\dagger_{jb})
     (U_{bl}U^\dagger_{ka})\frac{1}{\l_a-\l_b}\frac{\der}{\der\l_a}
     \nonumber \\
     & & +{\cal{O}}(\frac{\der}{\der U}).
 \label{derid} \err
We work with a restricted Hilbert space consisting only of states
which are annihilated by $\der/\der U$.
We therefore disregard
the last term in (\ref{derid}).  Note that (\ref{teed})
is recovered when (\ref{derid}) is acted on with
$\sum_l\sum_k\d_{il}\d_{jk}$, as expected.
It is useful to define a function
\bq w=-\sum_i\sum_{j\ne i}\ln|\l_i-\l_j|
 \eq
which has the following properties,
\bq \frac{\der w}{\der\l_i} = -\sum_{j\ne i}\frac{1}{\l_i-\l_j}
 \eq
and
\brr \frac{\der^2 w}{\der\l_m\der\l_n}=
     \left\{\begin{array}{ll}
            \sum_{k\ne i}{1}/{(\l_i-\l_k)^2} & \mbox{$;  m=n$} \\
            -1/(\l_m-\l_n)^2 & \mbox{$;m\ne n$}
      \end{array}\right. ,\label{zil}
 \label{derww} \err
The first two terms in (\ref{hambone}) are identical to the
Hamiltonian treated in Appendix A.  To connect with the notation
used in Appendix A, we define
\bq V(M)=\hf Tr (\frac{\der W(M)}{\der{M}})^2.
 \label{vdef} \eq
The result, (\ref{trombone}), is directly applicable.  Using
(\ref{derww}) and (\ref{vdef}), and noting that
$\hat{\Pi}_{\l_i}=-i\der/\der\l_i$, (\ref{trombone})
and, hence, the first two terms of (\ref{hambone})
can be written as follows,
\brr & & \sum_{ij}\Biggl\{\hf\hat{\Pi}_{M_{ij}}\hat{\Pi}_{M_{ji}}
     +\hf\frac{\der W(M)}{\der M_{ij}}\frac{\der W(M)}{\der M_{ji}}\Biggr\}
     \nonumber \\
     & &=\sum_i\Biggl\{\hf\hat{\Pi}_{\l_i}^2
     +i\frac{\der w}{\der\l_i}\hat{\Pi}_{\l_i}
     +\hf(\frac{\der W(\l_i)}{\der\l_i})^2\Biggr\}.
 \label{sface} \err
We now concentrate on the last term in (\ref{hambone}).
Using the relation (\ref{rel2}) it is easily seen that
\brr & & \sum_{ijkl}[\bar{\Psi}_{ij},\Psi_{kl}]
     \frac{\der^2W(M)}{\der M_{ij}\der M_{kl}} \nonumber \\
     & & =-\hf\sum_{ij}\frac{\der^2W(M)}{\der M_{ij}\der M_{ji}}
     +\sum_{ijkl}\bar{\Psi}_{ij}\Psi_{kl}
     \frac{\der^2W(M)}{\der M_{ij}\der M_{kl}}.
 \err
Now, using (\ref{derid}) and (\ref{chidef}) it is straightforward
to show the following,
\brr \sum_{ij}\frac{\der^2W(M)}{\der M_{ij}\der M_{ji}}
     &=& \sum_i\Biggl\{\frac{\der^2W(\l)}{\der\l_i^2}
     -2\frac{\der w(\l)}{\der\l_i}\frac{\der W(\l)}{\der\l_i}\Biggr\}
     \nonumber \\
     \sum_{ijkl}\bar{\Psi}_{ij}\Psi_{lk}
     \frac{\der^2W(M)}{\der M_{ij}\der M_{kl}}
     &=& \sum_{ij}\bar{\chi}_{ii}\chi_{jj}
     \frac{\der^2W(\l)}{\der\l_i}{\der\l_j} \nonumber \\
     & & +2\sum_i\sum_{j\ne i}\bar{\chi}_{ij}\chi_{ji}
     \frac{1}{\l_i-\l_j}\frac{\der W(\l)}{\der\l_i}.
 \label{quid}\err
We now further restrict the Hilbert space to include only those
states $|S>$ which are annihilated by ``off-diagonal" fermionic
creation operators,
$\chi_{ij}$, where $i\ne j$.  The last term in (\ref{quid})
then annihilates this subspace of states and can be neglected.
We abbreviate the
diagonal fermions, $\chi_{ii}$, by denoting them $\chi_i$.
Using the quantization condition, $\{\bar{\chi}_i,\chi_j\}=\d_{ij}$,
it is now straightforward to show that
\brr & & \hf\sum_{ijkl}[\bar{\Psi}_{ij},\Psi_{kl}]
     \frac{\der^2W(M)}{\der M_{ij}\der M_{kl}} \nonumber \\
     & & \hspace{.3in}
     \sum_i\frac{\der w(\l)}{\der\l_i}\frac{\der W(\l)}{\der\l_i}\}
     + \hf\sum_{ij}[\bar{\chi}_i,\chi_j]
     \frac{\der^2W(\l)}{\der\l_i\der\l_j}.
 \err
Immediately, by combining (B.1) and (B.12), we find that
\brr \hat{H}_S
     &=& \hf\hat{\Pi}_{\l_i}^2
     +i\frac{\der w}{\der\l_i}\hat{\Pi}_{\l_i}
     +\hf(\frac{\der W}{\der\l_i})^2
     +\frac{\der w}{\der\l_i}\frac{\der W}{\der\l_i}\} \nonumber \\
     & & \hspace{.5in}
     + \hf\sum_{ij}[\bar{\chi}_i,\chi_j]
     \frac{\der^2W(\l)}{\der\l_i\der\l_j},
 \label{hammy} \err
where the subscript $S$ indicates the restriction to the singlet,
fermion diagonal subspace of states.
Over the singlet sector, the partition function is
\brr Z_N(b_n) &=& \int [d\Pi_\l][d\l][d\chi][d\bar{\chi}]
    \exp i\int dt
    \sum_{ij}\bar{\chi}_i\chi_j\frac{\der^2 w}{\der\l_i\der\l_j}
    \nonumber \\
    & & \hspace{.5in}
    \times\exp i\int dt
    \sum_i\Biggl\{\Pi_{\l_i}\dot{\l}_i-i\bar{\chi}_i\dot{\chi}_i-H_S\Biggr\}
 \err
In this expression, the first exponential factor is a
Jacobian associated with the parameterization of the fermion
fields.  It is necessary because the measure on the Hilbert
space becomes nontrivial when we restrict to the ``diagonal"
states, $\chi_i$, which is essentially a choice of curvilinear
coordinates in functional space.
Inserting (\ref{hammy}) and rearranging, this partition function
can be expressed as
\brr Z_N(b_n) &=& \int [d\Pi_{\l}][d\l][d\chi][d\bar{\chi}]
     \nonumber \\
     & & \hspace{.2in}
     \times\exp i\int dt\sum_i
     \Biggl\{-\hf\Pi_{\l_i}^2+(\dot{\l}_i+i\frac{\der w}{\der\l_i})
     \Pi_{\l_i}\} \nonumber \\
     & & \hspace{.2in}
     \times\exp i\int dt
     \{\sum_i[-\hf(\frac{\der W}{\der\l_i})^2
     -\frac{\der w}{\der\l_i}\frac{\der W}{\der\l_i}
     -i\bar{\chi}_i\dot{\chi}_i] \nonumber \\
     & & \hspace{.6in}
     +\sum_{ij}\bar{\chi}_i\chi_j\frac{\der^2(W+w)}{\der\l_i\der\l_j}\Biggr\}.
 \err
The Gaussian $[d\Pi_\l]$ integration is straightforward and the details
are the same as those described in Appendix A.  Performing the
$[d\Pi_\l]$ integration, it is readily found that
\bq Z_N(b_n)=\int [d\l][d\chi][d\bar{\chi}]
    \exp i\int dt L_S,
 \eq
where
\bq L_S = \sum_i\Biggl\{
     \hf\dot{\l}_i^2
     -\hf(\frac{\der(W+w)}{\der\l_i})^2
     -\frac{i}{2}(\bar{\chi}_i\dot{\chi}_i-\dot{\bar{\chi}}_i\chi_i)\Biggr\}
     -\sum_{ij}\bar{\chi}_i\chi_j\frac{\der^2(W+w)}{\der\l_i\der\l_j}.
 \eq
This is the result cited in section 3.

\renewcommand{\theequation}{C.\arabic{equation}}
\setcounter{equation}{0}
{\fl{\bf Appendix C: Proper Implementation of the Collective Field
Constraint Condition}}

In this Appendix we discuss a technical issue associated with the
proper implementation of constraints when constructing the
high density collective field theory.  This issue is relevant to
subsections 4.3 and 4.4.  The field equations shown in
(\ref{mott}) were derived from the Lagrangian
(\ref{lagcollnoncan}) with $\Lambda=0$ by the usual variation
method.  At the end of subsection 4.3 we made the assertion that
this procedure gives rise to the correct canonical theory and could
by used with impunity.  We proceed to prove this.

It follows from the definitions (\ref{colldef}), that
the field $\vphi$ in equation (\ref{lagcollnoncan})
must satisfy the following constraint equation,
\bq \int \vphi'(x)dx=N,
 \label{constable} \eq
even when the large
$N$ limit is taken.  It can be shown that this is the only
constraint which the high density fields are required to satisfy.
When varying the Lagrangian (\ref{lagcollnoncan}) to derive
field equations, this constraint must be accounted for.
A powerful way to implement the constraint is to
amend the Lagrangian by the addition of a Lagrange
multiplier term. In this case, the purely bosonic part of the
collective field Lagrangian becomes
\bq L_B = \int dx\Biggl\{\frac{\dot{\vphi}^2}{2\vphi'}
     -\frac{\pi^2}{6}\vphi^{'3}
     +\hf(\w^2x^2-\Lambda)\vphi'
     +\mu(\vphi'-\frac{N}{L})\Biggr\},
 \label{lagmod} \eq
where $\mu$ is the Lagrange multiplier.
The fermionic parts of the Lagrangian and the fermionic field equations
are unnaffected by this concern and we omit them from this
discussion.  We then vary the Lagrangian (\ref{lagmod})
with respect to both $\vphi$, which is now unconstrained,
and to $\mu$.  This gives us a
coupled system of equations which determine both the stable
configuration, $\tilde{\vphi}_0(x)$, which will depend on $\mu$,
as well as a relation between $\mu, N$ and $L$.
Doing this, we find
the $\vphi$ equation and the $\mu$ equation, respectively, to be
\brr   \der_t(\frac{\dot{\vphi}}{\vphi'})
       -\hf\der_x(\frac{\dot{\vphi}^2}{\vphi^{'2}}+\pi^2\vphi^{'2}
       -\w^2x^2+\Lambda-2\mu) &=& 0 \nonumber \\
       \int dx\vphi'(x) &=& N.
 \label{prancer} \err
The first of these equations is solved, for the static case
$\dot{\tilde{\vphi}}_0=0$, by the following expression,
\bq \tilde{\vphi}_0'=\frac{1}{\pi}\sqrt{\w^2x^2-\frac{1}{{\rm g}}},
 \label{solphi} \eq
where $\frac{1}{{\rm g}}=2\mu-\Lambda+C$ and $C$ is an arbitrary
integration constant.  There are two possibilities.  Either
${\rm g}>0$ or ${\rm g}<0$.  We will consider
each of these cases independently.

a) ${\rm g}>0$:  In this case,
$\tilde{\vphi}_0'$ is only defined for
$(\w\sqrt{{\rm g}})^{-1}\le|x|\le L/2$.  The second equation in
(\ref{prancer}) then requires that
\bq \frac{2}{\pi}\int_{(\w\sqrt{{\rm g}})^{-1}}^{L/2}dx
    \sqrt{\w^2x^2-\frac{1}{{\rm g}}}=N.
 \eq
Integrating and performing some algebra, this equation becomes
\bq N=\frac{1}{2\pi\w}
    \Biggl\{\hf L^2\w^2-\frac{1}{{\rm g}}\ln (\frac{L^2\w^2}{{\rm g}})
-\frac{1}{{\rm g}}\Biggr\} +{\O}(\frac{1}{L^2}).
 \eq

b) ${\rm g}<0$:  In this case,
$\tilde{\vphi}_0'$ is defined for all
$|x|\le L/2$.  The second equation of (\ref{prancer}) then
requires that
\bq \frac{2}{\pi}\int_0^{L/2}dx
    \sqrt{\w^2x^2-\frac{1}{{\rm g}}}=N.
 \eq
Integrating and performing some algebra, this equation becomes
\bq N=\frac{1}{2\pi\w}
    \Biggl\{\hf L^2\w^2-\frac{1}{{\rm g}}\ln (\frac{L^2\w^2}{-{\rm g}})
-\frac{1}{{\rm g}}\Biggr\} + {\O}(\frac{1}{L^2}).
 \label{rudolph2} \eq
Combining the above results
we see that, regardless
of the sign of ${\rm g}$, the constraint equation (\ref{constable})
is embodied completely in the following relation,
\bq N=\frac{1}{2\pi\w}
    \Biggl\{\hf L^2\w^2-\frac{1}{{\rm g}}\ln (\frac{L^2\w^2}{|{\rm g}|})
    -\frac{1}{{\rm g}}\Biggr\}+\cal{O}(\frac{1}{L^2}).
 \label{rover} \eq

Now, recall that it was necessary to specify the $N$ dependence
of the coefficients of the superpotential for large $N$.  One result of
this is that $\Lambda=\hf Nc_1^2$.  In much the same manner,
we must now specify the large $N$ behavior of the new coefficients
$\mu$ and $C$.  The appropriate choice is
\bq 2\mu+C=\hf Nc_1^2.
 \eq
It follows that, for large $N$, $\frac{1}{g}$ is a finite constant.
With this in mind, let us analyze
(\ref{rover}) in the limit of large $N$ and large $L$.
Since $\frac{1}{g}$ is a constant in this limit, it is clear that
this equation simplifies to
\bq N=\frac{\w L^2}{4\pi}.
 \label{nvsl} \eq
That is, not only do we take the $N\rightarrow\infty,
L\rightarrow\infty$ limit but we must do so in such a way that
(\ref{nvsl}) is satisfied.  Note that, since
$N/L\propto L\rightarrow\infty$, this condition is compatible
with the high density eigenvalue condition.  The result of all this is,
in fact, very simple.  It implies that (\ref{solphi}), with an arbitrary
constant $\frac{1}{g}$, is the solution
of the $\vphi$ equation of motion in the appropriate
$N\rightarrow\infty,
L\rightarrow\infty$ limit.  Note that this is exactly the result
that would have been obtained from the high density Lagrangian
if we simply took $\Lambda=0$, treated $\vphi$ as an unconstrained
field and ignored the question of the constraint
(\ref{constraint}).  In this case $\frac{1}{g}$ would arise as an
arbitrary integration constant.  This justifies the statements made
in subsection 4.3.

\end{document}